\documentclass[fleqn,usenatbib]{mnras}
\usepackage{newtxtext,newtxmath}
\usepackage[T1]{fontenc}
\usepackage[normalem]{ulem}
\usepackage{mathtools}
\usepackage{cuted}

\title[Cliff collapse on Comet 67P]{Cliff collapse on Comet 67P/Churyumov--Gerasimenko -- II. Imhotep and Hathor}

\author[Bj\"{o}rn J. R. Davidsson]{
Bj\"{o}rn J. R. Davidsson,$^{1}$\thanks{E-mail: bjorn.davidsson@jpl.nasa.gov}
\\
$^{1}$Jet Propulsion Laboratory, California Institute of Technology,  M/S 183--601, 4800 Oak Grove Drive, Pasadena, CA 91109, USA
}

\date{Accepted 2024 February 29. Received 2024 February 01; in original form 2023 September 21}

\pubyear{2024}

\begin{document}
\label{firstpage}
\pagerange{\pageref{firstpage}--\pageref{lastpage}}
\maketitle

%
\begin{abstract}
Cliff collapses on Comet 67P/Churyumov--Gerasimenko expose relatively pristine nucleus matter and offer rare opportunities to characterise ice--rich 
comet material. Here, Microwave Instrument for \emph{Rosetta} Orbiter (MIRO) observations of two collapsed or crumbling cliffs in the Imhotep and Hathor regions have been assembled. 
The empirical diurnal antenna temperature curves are analysed with thermophysical and radiative transfer models in order to place constraints on the physical 
properties and degrees of stratification in the near--surface material. The Imhotep site consists of an exposed dust/water--ice mixture with thermal inertia 
100--$160\,\mathrm{J\,m^{-2}\,K^{-1}\,s^{-1/2}}$, having sublimating $\mathrm{CO_2}$ ice located $11\pm 4\,\mathrm{cm}$ below the surface. Its estimated age is consistent with an outburst 
observed on 2014 April 27--30. The Hathor site has a $0.8\pm 0.2\,\mathrm{cm}$ dust mantle, a thermal inertia of $40\pm 20\,\mathrm{J\,m^{-2}\,K^{-1}\,s^{-1/2}}$, no 
$\mathrm{CO_2}$ ice to within $1.0\,\mathrm{m}$ depth, and a mantle bulk density of $340\pm 80\,\mathrm{kg\,m^{-3}}$ that is higher than the theoretically expected 
$180\pm 10\,\mathrm{kg\,m^{-3}}$, suggesting that compression has taken place. 
\end{abstract}


\section{Introduction} \label{sec_intro}

One of the biggest surprises during the \emph{Rosetta} mission \citep{glassmeieretal07,tayloretal17} to Comet 67P/Churyumov--Gerasimenko (hereafter, 67P) was 
the discovery of pervasive, concentric layering of each nucleus lobe \citep{massironietal15,penasaetal17}. Due to differential erosion rates, these layers are manifested on the surface as 
intricate staircase systems of cliffs and plateaus \citep{sierksetal15,thomasetal15a} with spectrophotometrically peculiar properties \citep{ferrarietal18,tognonetal19,davidssonetal22c}. 
The cliffs are major sources of activity and producers of coma jet--like features \citep{vincentetal16,vincentetal17}. The frequently occurring talus cones and boulders underneath cliffs \citep{pajolaetal15} 
show that mass wasting, ranging from small--scale crumbling to large--scale collapse of $100\,\mathrm{m}$--sized wall segments \citep{pajolaetal17}, is an important part of comet nucleus evolution. 

The collapse of cliff walls suddenly exposes deep and relatively pristine nucleus material that has been largely shielded from solar heating by the significantly more processed 
dust mantle material that is dark, ice--poor, and dominated by refractory silicates and organics \citep{fornasieretal15,capaccionietal15,quiricoetal16,mennellaetal20}. 
Collapses provide rare opportunities to study the icy component of comet material, and to characterise physical properties that are more representative of the nucleus bulk than the 
surface. Developing a better understanding of comet nucleus interiors is crucially important in order to understand their formation and evolution, and by extension, constraining the 
properties of the Solar nebula in which they formed. 

\citet[][hereafter, Paper~I]{davidsson24} analysed observations by the microwave instrument \emph{Rosetta}/MIRO \citep{gulkisetal07} of the Aswan cliff before and after its collapse. 
By forward--modelling the thermophysics of specific nucleus models, and feeding the resulting temperature profiles to a radiative transfer solver that generates synthetic antenna 
temperature curves, it is possible to constrain the composition, stratification, thermal inertia, diffusivity, extinction coefficients, and single--scattering albedos of the collapse sites if and 
when those curves are matching the MIRO observations. The current paper extends this type of analysis to two other sites, located in Imhotep and in Hathor \citep[for nucleus regional names 
and definitions, see][]{thomasetal18}.

\begin{figure*}
\centering
\begin{tabular}{cc}
\scalebox{0.3}{\includegraphics{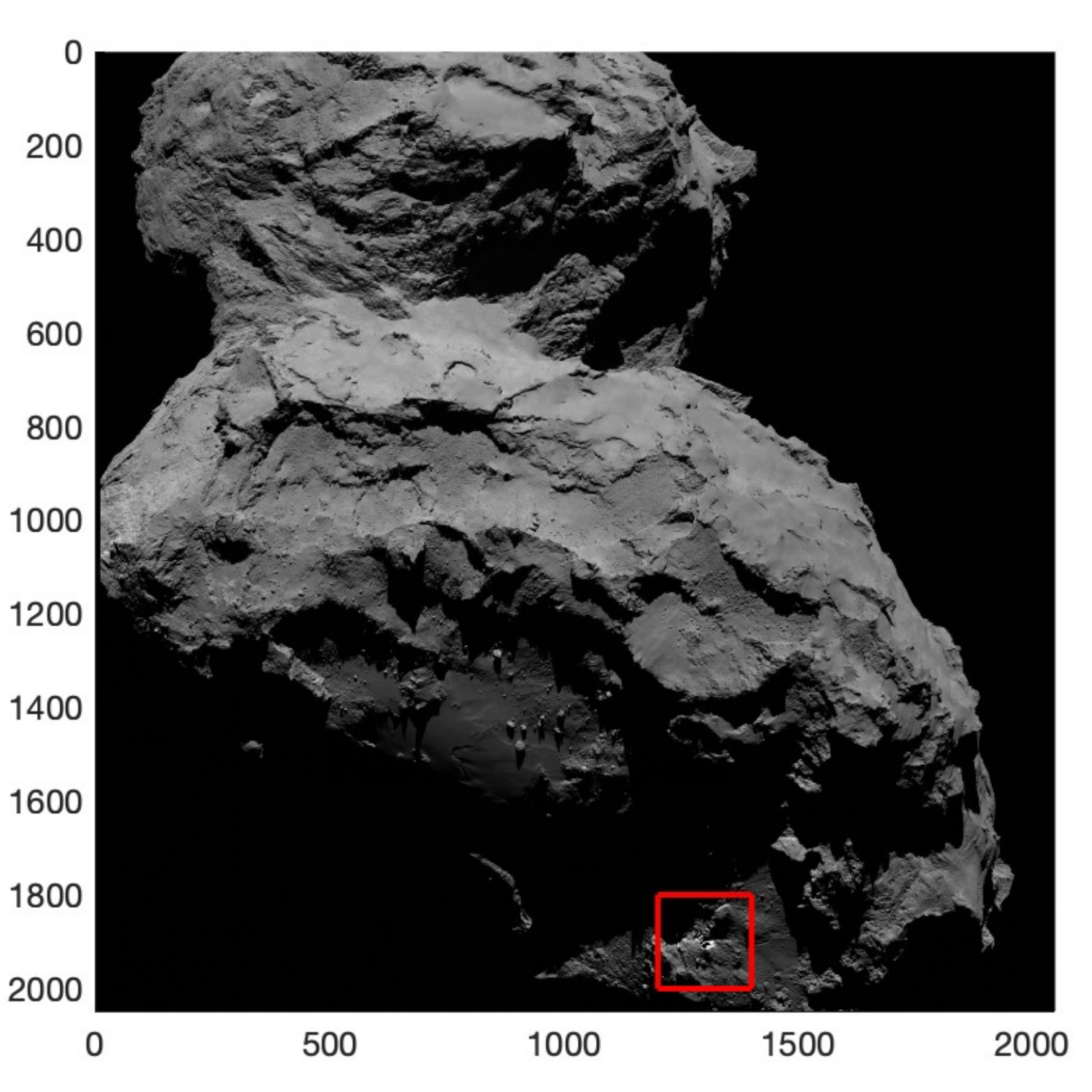}} & \scalebox{0.3}{\includegraphics{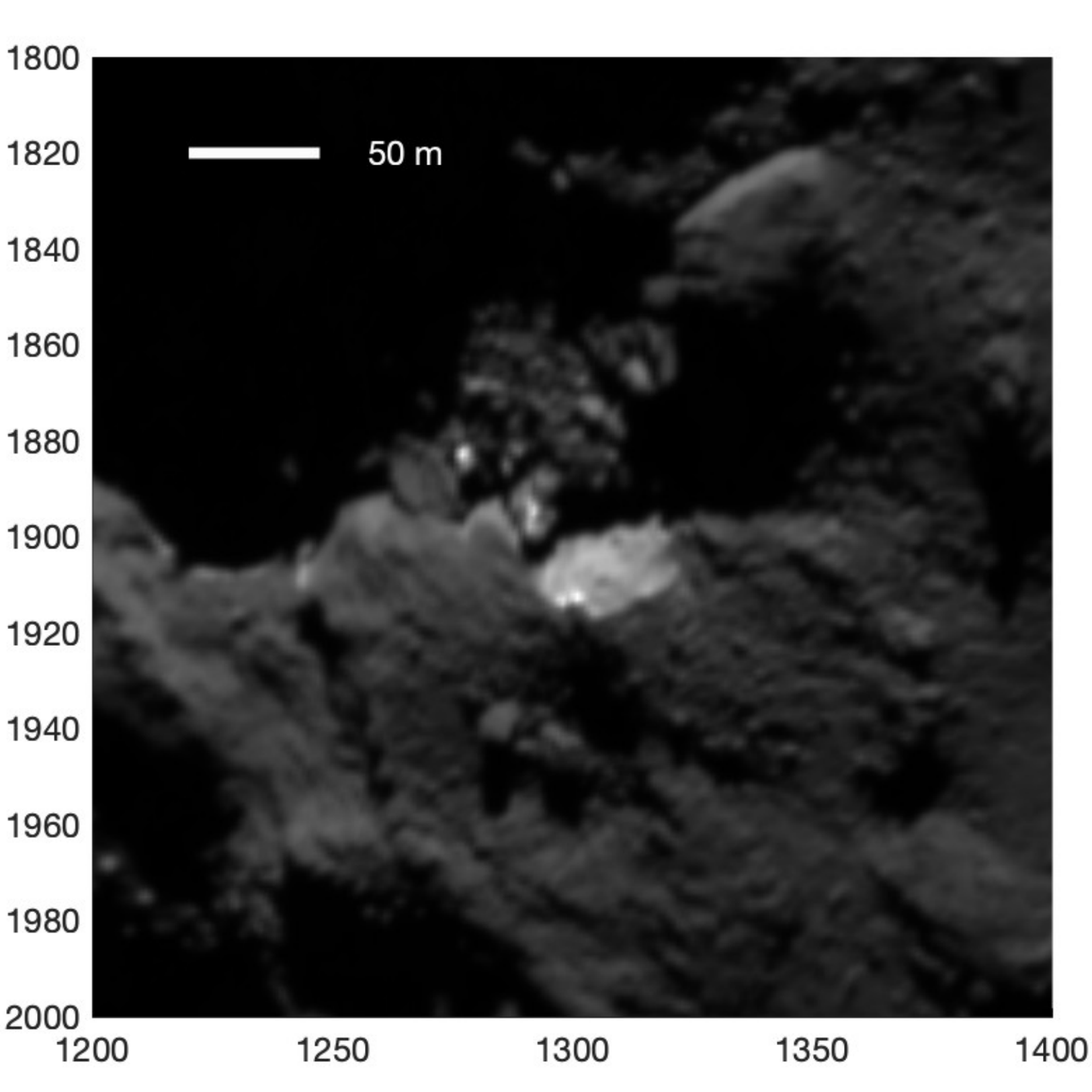}}\\
\end{tabular}
     \caption{\emph{Left:} Context image of the nucleus of Comet 67P showing the Imhotep collapse site within the red square. It is located on the large lobe, while the small 
lobe is visible in the top third of the panel. \emph{Right:} This is a zoom on the red square from the left panel. The Imhotep collapse site clearly stands out as a 
brighter region. Both panels show image MTP6/n20140816t165914570id4ff22.img \protect\citep{sierksetal20a} that was acquired on 2014 August 16 with $\sim 1.8\,\mathrm{m\,px^{-1}}$ resolution when \emph{Rosetta} was $103\,\mathrm{km}$ from the comet. Axis labels show pixel ID numbers.}
     \label{fig_ImhotepA}
\end{figure*}

In Paper~I it was found that the pre--collapse Aswan site dust mantle had a thickness of at least $h_{\rm m}\geq 3\,\mathrm{cm}$ and a thermal inertia of $\Gamma\approx 30\,\mathrm{J\,m^{-2}\,K^{-1}\,s^{-1/2}}$ 
(hereafter MKS). The post--collapse material had an estimated dust/water--ice mass ratio of $\mu=0.9\pm 0.5$, a diffusivity of $\mathcal{D}=0.1\,\mathrm{m^2\,s^{-1}}$, a thermal inertia of $\Gamma=25\pm 15\,\mathrm{MKS}$, 
and the observations were consistent with the independently estimated carbon dioxide molar abundance $\mathrm{CO_2/H_2O}=0.32$ relative to water \citep{davidssonetal22}, and a nucleus bulk 
density of $\rho_{\rm bulk}=535\,\mathrm{kg\,m^{-3}}$ \citep{jordaetal16,preuskeretal17,patzoldetal19}. A thin ($h_{\rm m}\leq 3\,\mathrm{mm}$) dust mantle developed $\sim 7$ months 
after the collapse, having a thermal inertia in the $\Gamma=10$--$45\,\mathrm{MKS}$ range. The sublimation front depth of the $\mathrm{CO_2}$  supervolatile (sv) was 
$h_{\rm sv}=0.4\pm 0.2\,\mathrm{cm}$ after 5 months, $h_{\rm sv}=2.0\pm 0.3\,\mathrm{cm}$ after 7 months, and $h_{\rm sv}=20\pm 6\,\mathrm{cm}$ after 11 months. In the current 
paper, attempts are made to place similar constraints on the material at the Imhotep and Hathor collapse sites. 

\begin{figure}
\scalebox{0.3}{\includegraphics{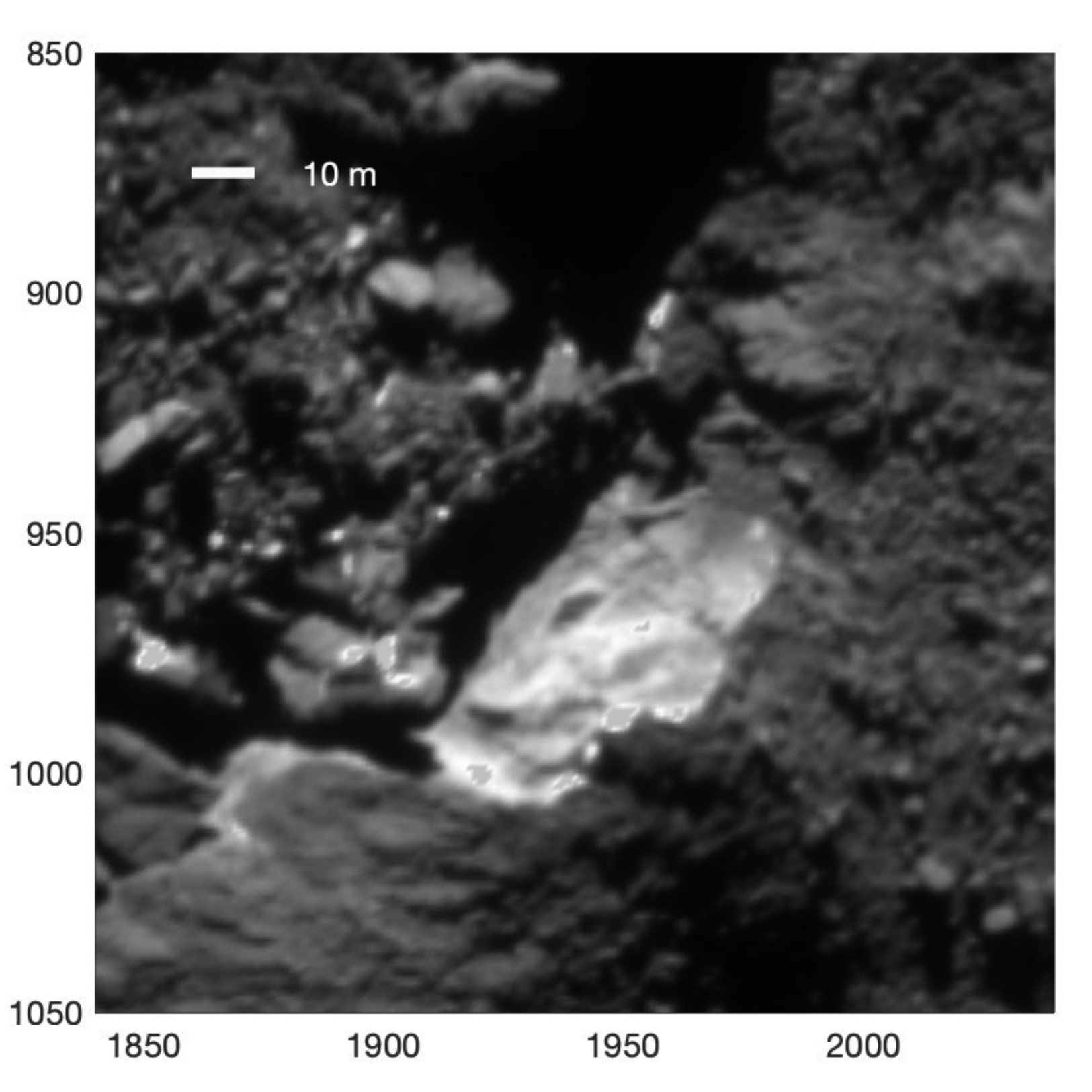}}
     \caption{This image shows the Imhotep collapse site, at somewhat higher resolution than in Fig.~\ref{fig_ImhotepA}. To brighten the overall image and enhance visibility, the radiance factors of a 
few unusually bright pixels have been re-set to lower values. They appear as small grey patches within whitish terrain and are artificial. Image MTP7/n20140905t080224552id4ff71.img \protect\citep{sierksetal20c} that was acquired on 2014 September 5 with $\sim 0.76\,\mathrm{m\,px^{-1}}$ resolution when \emph{Rosetta} was $42.8\,\mathrm{km}$ from the comet.}
     \label{fig_ImhotepB}
\end{figure}

The Imhotep collapse site is seen in \emph{Rosetta}/OSIRIS \citep{kelleretal07} images shown in Fig.~\ref{fig_ImhotepA} (all OSIRIS images are available at the NASA Planetary Data System, 
PDS\footnote{https://pds--smallbodies.astro.umd.edu/data\_sb/missions/rosetta/\\index\_OSIRIS.shtml}). The context image (left panel) is dominated by the large lobe, 
centrally showing the large smooth--terrain plain of Imhotep. At its border, where the landscape topography becomes increasingly complex, lays the cliff wall in question. This collapse site, 
easily recognisable by its higher albedo (right panel) was already present when \emph{Rosetta} arrived (in fact, this image was acquired less than two weeks after arrival). Figure~\ref{fig_ImhotepB} shows 
another view, at somewhat better resolution. The bright scar appears rather featureless at $\sim 0.76\,\mathrm{m\,px^{-1}}$ resolution, in contrast to the rougher and darker terrains surrounding it.

For RGB colour--composite images of the Imhotep collapse region, see Fig.~7B in \citet{augeretal15a} and Fig.~8d in \citet{pommeroletal15}. Such images reveal that 
the bright regions are spectrally bluer than their surroundings. Cross--comparison between OSIRIS spectrophotometry and \emph{Rosetta}/VIRTIS \citep{coradinietal07} spectroscopy 
show that bluish terrain displays the characteristic $\sim 3\,\mathrm{\mu m}$ absorption feature of water ice \citep{baruccietal16}. It is therefore evident that water ice is abundant at 
the collapse site, while it is virtually absent in the surrounding dust mantle material.

\begin{figure*}
\centering
\begin{tabular}{cc}
\scalebox{0.3}{\includegraphics{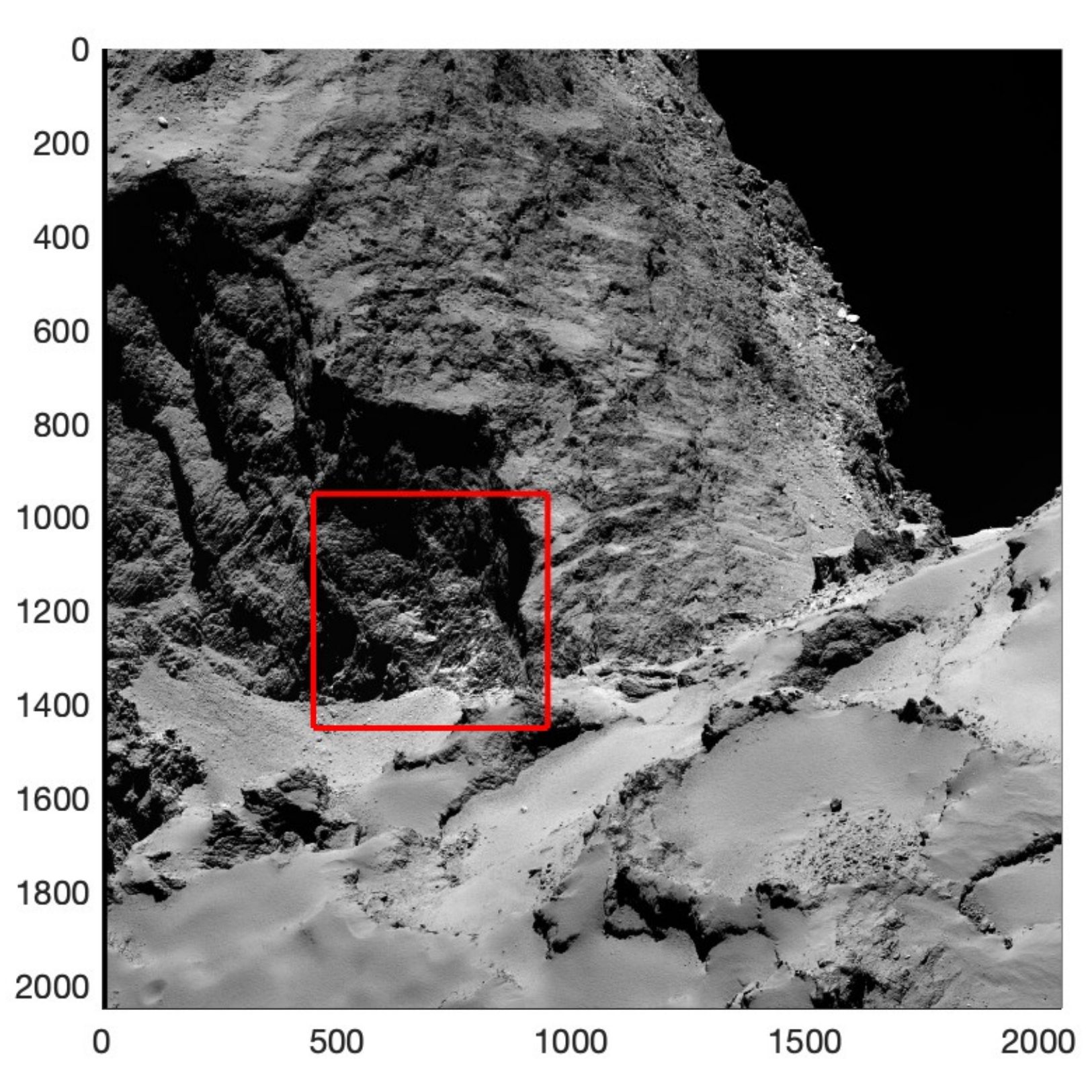}} & \scalebox{0.3}{\includegraphics{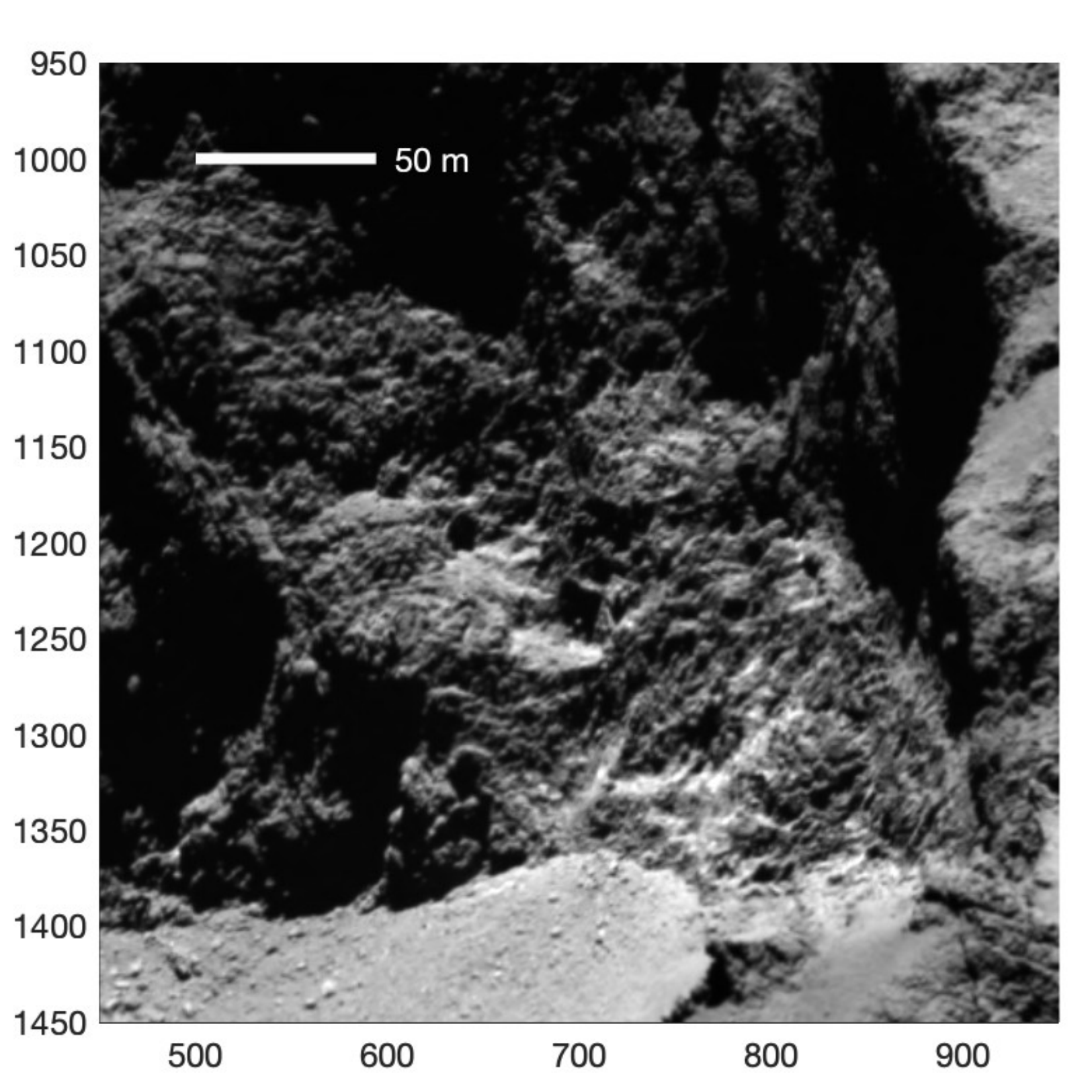}}\\
\end{tabular}
     \caption{\emph{Left:} Context image of the nucleus of Comet 67P showing the Hathor alcove within the red square. The sharp boundary to Anuket runs parallel just 
within the right side of the square. The alcove is located on the small lobe. A small fraction of the large lobe is visible in the foreground in the lower third of the panel. 
Note the morphological differences between these parts of the small and large lobes, being dominated by consolidated and smooth terrains, respectively. \emph{Right:} This is a zoom on the red square 
from the left panel. Spectrophotometry reveals unusually bright and bluish material at and beneath the Hathor alcove, though much of its surface is covered by dark dust. Both panels show 
image MTP7/n20140912t195732364id4ff22.img \protect\citep{sierksetal20c} that was acquired on 2014 September 12 with $\sim 0.53\,\mathrm{m\,px^{-1}}$ resolution when \emph{Rosetta} was $29.9\,\mathrm{km}$ from the comet.}
     \label{fig_Hathor}
\end{figure*}

Spectrophotometrically bluish terrain is also found elsewhere. One particular region that attracted attention early in the \emph{Rosetta} mission was the 
part of Hathor dubbed the `alcove' by  \citet{thomasetal15a}. An RGB colour--composite image of the alcove is seen in Fig.~5 (left) of that paper. 
The region is also shown here in Fig.~\ref{fig_Hathor}. \citet{thomasetal15a} describe $\leq 10\,\mathrm{m}$ bright spots on the cliff wall, and additionally mention 
2--$5\,\mathrm{m}$ spots in the underlying talus that are 20 per cent brighter and bluer than the surroundings. It therefore seems like this cliff wall was actively 
experiencing small--scale crumbling in 2014 August, that exposed ice--rich material on the wall and deposited ice--rich debris on the valley floor underneath. 
However, it is clear that the majority of the surface area is covered by dark and refractory material. This could mean that it once resembled the more recently collapsed 
Aswan and Imhotep cliffs, but now has aged to the point that the dust mantle largely has recovered its pre--collapse appearance. Alternatively, the alcove represent another 
type of wall on 67P that is not prone to large--scale collapse, but merely fragments one $10\,\mathrm{m}$--block at a time. However, it is perhaps relevant that the Hathor 
alcove lies at a peculiar boundary to the region Anuket. The boundary is very sharp and gives the impression of a brittle material from which a big chunk has been broken 
off along a distinct crack, thus revealing the alcove that is stratigraphically interior to Anuket \citep{elmaarryetal15}. However, there is no way of knowing if that was a 
one--time big event, or a long series of much smaller events.

\section{Methodology} \label{sec_method}

The overall approach, and the numerical models being employed here, are very similar to those of \citet{davidssonetal22b} and Paper~I. 
Therefore, the information provided here is very brief in the interest of saving space, and the reader is strongly encouraged to 
consult the previously mentioned publications for further details. 

The first step is to identify the two regions on the 67P shape model. This is necessary both for being able to calculate illumination 
conditions that are specific to the regions, and to facilitate searches in the MIRO database for suitable observations. The nucleus shape model 
SHAP5 version 1.5 \citep{jordaetal16}, is here degraded from $3.1\cdot 10^6$ to $5\cdot 10^4$ facets. The regions are located through visual 
inspection and cross--comparison of images and the shape model. 

The Imhotep site is represented by 7 facets, with a largest mutual separation of $66\,\mathrm{m}$ (providing a measure of the size of the region). 
The number of `terrain facets' (having the geometric capability of shadowing and self--heating the Imhotep site) is 1,404 and the number of 
`surrounding facets' (being capable of shadowing the terrain facets) are 10,629, which means that 12,033 facets were included in the 
illumination condition calculations. The Hathor site is represented by 20 facets,  with a largest mutual separation of $105\,\mathrm{m}$. 
The number of terrain facets is 2,509. The number of surrounding facets is 14,004, so that a total of 16,513 facets were included in the 
illumination condition calculation. 

The model by \citet{davidssonandrickman14} is used to calculate illumination conditions for Imhotep and Hathor. These calculations take 
the nucleus shape, topography,  spin axis orientation, and the rotational phase (including corrections due to the time--dependent rotational period of the comet) 
into account during orbital motion, when calculating the direct solar flux onto a given Imhotep or Hathor facet. The equilibrium temperature is calculated for 
illuminated terrain facets, and parts of their thermal reradiation illuminates the facets at the Imhotep and Hathor sites according to the mutual view factors. 
The surrounding facets are used to calculate which terrain facets are being in shadow (if so their self--heating contributions are switched off).

The illumination conditions at each of the 7 Imhotep facets and 20 Hathor facets are evaluated every $20\,\mathrm{min}$ from the 2012 May 23 aphelion 
to the end of 2015 January (no MIRO data beyond that date was included in this study). Because illumination conditions change slowly with time, and because 
these calculations are very time--consuming, the actual conditions were evaluated every $12^{\rm th}$ nucleus rotation and copied to the following 11 ones. 
At a given nucleus rotational phase, the area--weighted mean fluxes were calculated for the 7 Imhotep and 20 Hathor facets, respectively. These mean 
illumination conditions were fed to the thermophysical model. 

The thermophysical model employed here is called the `Numerical Icy Minor Body evolUtion Simulator' or \textsc{nimbus}. It is fully described 
by \citet{davidsson21}, but also see \citet{davidssonetal22b} and Paper~I for usage specific to MIRO data analysis. \textsc{nimbus} is here used to 
study porous mixtures consisting of refractories, hexagonal (crystalline) water ice, and $\mathrm{CO_2}$. Other capabilities of \textsc{nimbus}, including 
amorphous ice and release of trapped $\mathrm{CO}$ and $\mathrm{CO_2}$ during crystallisation into cubic water ice, the cubic--hexagonal transition and 
associated $\mathrm{CO}$ and $\mathrm{CO_2}$ release, segregation of $\mathrm{CO_2:CO}$ mixtures, and the sublimation of CO ice, are not utilised here. 
\textsc{nimbus} solves a coupled system of differential equations describing the conservation of energy and masses of ices and vapours. These equations 
describe the solid--state and radiative transport of heat, the energy consumption during $\mathrm{H_2O}$ and $\mathrm{CO_2}$ sublimation, the transport 
of energy (advection) and vapour mass during gas diffusion within the porous medium (along temperature and vapour partial pressure gradients), and energy 
release during recondensation of $\mathrm{H_2O}$ and $\mathrm{CO_2}$ vapour. Nominally, both heat transport and gas diffusion are calculated in 
two spatial dimensions (radially and latitudinally) for a spherical body with illumination conditions specific for each latitude during nucleus rotation and orbital motion. 
However, here the variant \textsc{nimbusd} is employed, which sacrifices latitudinal mass and energy transport (which are not important for the relatively 
short time scales considered here) in order to enable erosion of the upper surface in response to outgassing. In any case, each latitude is here fed with the 
same Imhotep or Hathor illumination conditions. The ices are considered finite resources, which 
means that water ice withdraws gradually and leaves behind a dust mantle that only consists of dust (unless vapours recondenses temporarily during night time). 
The $\mathrm{CO_2}$ withdraws as well, resulting in a stratified near--surface region that contains a layer between the dust mantle and the $\mathrm{CO_2}$ 
sublimation front that only contains refractories and water ice. The porosity of the nucleus evolves as a result of ice sublimation and vapour recondensation, which 
affects both the effective heat conductivity, heat capacity, and diffusivity of the medium. In case outgassing is so strong that the entire dust mantle erodes away, 
the $\mathrm{CO_2}$ activity will start to erode the dust and water ice mixture as well. All expressions for heat conductivity of compacted substances, specific 
heat capacities, latent heats, and vapour saturation pressures are temperature--dependent and species--specific, and are based on expressions established in 
laboratory measurements. In short, \textsc{nimbus} is a state--of--the--art thermophysics code that includes all processes that typically are considered important 
for comet activity in the thermophysics literature. 

The final outcome of a \textsc{nimbus} simulation of largest importance here is the nucleus temperature as function of depth and time. These temperature 
profiles are passed to a radiative transport code called \textsc{themis}, described fully by \citet{davidssonetal22b}, but also see Paper~I. For a given 
temperature profile, \textsc{themis} calculates the radiance of emitted radiation as function of emergence angle. The code is evaluated for the 
specific wavelengths at which MIRO is observing -- the millimetre (MM) channel at wavelength $\lambda=1.594\,\mathrm{mm}$ and the submillimetre (SMM) 
channel at $\lambda=0.533\,\mathrm{mm}$. The code uses wavelength--specific extinction coefficients ($E_{\rm MM}$ and $E_{\rm SMM}$) and single--scattering 
albedos ($w_{\rm MM}=0$ in this type of problems, but $E_{\rm SMM}>0$ is employed when required). The radiances are recalculated to antenna temperatures, that 
can be directly compared with MIRO measurements. The final output of \textsc{themis} are diurnal MM and SMM antenna temperature curves, evaluated for 
the relevant emergence angle at each rotational phase according to the MIRO observational geometry.

Ideally, MIRO should have observed the Imhotep and Hathor collapse sites continuously throughout an entire nucleus rotation. However, such observations 
did not take place. Instead, the instrument observed the regions briefly at various times, separated by several days. Because the illumination conditions 
change very little with time during sufficiently short time intervals, and because the nucleus activity is highly regular \citep[as evident from the `clockwork repeatability of 
jets from one rotation to the next';][]{vincentetal16b}, MIRO observations acquired during a few weeks are therefore binned and time--shifted to a common `master period' in order 
to build empirical diurnal antenna temperature curves. Typically $\pm 2.5\,\mathrm{K}$ error bars are assigned to the bins. More detailed accounts of how the MIRO antenna temperature 
curves were constructed are given for the individual locations in sections~\ref{sec_results_Imhotep} and \ref{sec_results_Hathor}.

The MIRO observational archive \citep[PDS website\footnote{https://pds-smallbodies.astro.umd.edu/holdings/ro-c-miro-3-prl\\-67p-v3.0/dataset.shtml 
and \\https://pds-smallbodies.astro.umd.edu/holdings/ro-c-miro-3-esc1\\-67p-v3.0/dataset.shtml};][]{hofstadteretal18a,hofstadteretal18b} contains information 
about the beam centre interceptions with the nucleus surface (radial, latitudinal, and longitudinal coordinates) of each observation. Because these are known for the 
Imhotep and Hathor collapse site from the identification of relevant facets on the shape model, the archive can readily be searched for the relevant observations.

The quality of a given synthetic diurnal antenna temperature curve, with respect to the empirical one, is determined by calculating $\chi^2$ residuals 
and the incomplete gamma function value $Q$ \citep[see][and Paper~I]{pressetal02,davidssonetal22b}. Typically, the modelling is performed in stages, 
starting with refractories--only models, then moving on to dust$+\mathrm{H_2O}$ and finally dust$+\mathrm{H_2O}+\mathrm{CO_2}$ variants if no fits can be 
found for the more simple media. The parameter values of \textsc{nimbus} (e.~g., effective thermal inertia, diffusivity, composition, and degree of stratification) and 
\textsc{themis} (extinction coefficients and single--scattering albedos) are gradually estimated by generating a number of test cases, check how the synthetic antenna 
temperature curves changes in response to those parameter settings, and trying to tweak them until the model fits the data.

\section{Results} \label{sec_results}

\subsection{The Imhotep collapse site: 2014 Dec and 2015 Jan} \label{sec_results_Imhotep}

The Imhotep collapse site is $\sim 70\,\mathrm{m}$ across. The Full--Width Half--Maximum (FWHM) diameters of the 
MM and SMM beams are $207\,\mathrm{m}$ and $65\,\mathrm{m}$ at a nucleus distance of $30\,\mathrm{km}$, 
respectively \citep{schloerbetal15}. In order for the collapse site to dominate the radiation sensed by the beams, 
\emph{Rosetta} should ideally be closer than $30\,\mathrm{km}$ to the comet. During the mission this was predominantly the case 
during three periods: 1) 2014 September through 2015 January; 
2) the first half of 2016 March; 3)   2016 May through September. These periods were searched for MIRO observations of the 
Imhotep collapse site. Unfortunately, this region was rarely observed. The highest observational frequency was in 
2014 December with 21 data points, and in 2015 January with 53 data points. The data were averaged within $2.4\,\mathrm{min}$--wide bins 
(roughly corresponding to $1^{\circ}$ nucleus rotation). The bins were time--shifted to a common 
master period starting 2014 December 22, at $22:26:48\,\mathrm{UTC}$ (corresponding to daynumber\footnote{Daynumbers start with $d_{\rm n}=1$ on 2014 January 1, 00:00:00 UTC, increment 
daily by unity, and are used for MIRO time--keeping.} $d_{\rm n}=356.7$).

\begin{figure*}
\centering
\begin{tabular}{cc}
\scalebox{0.45}{\includegraphics{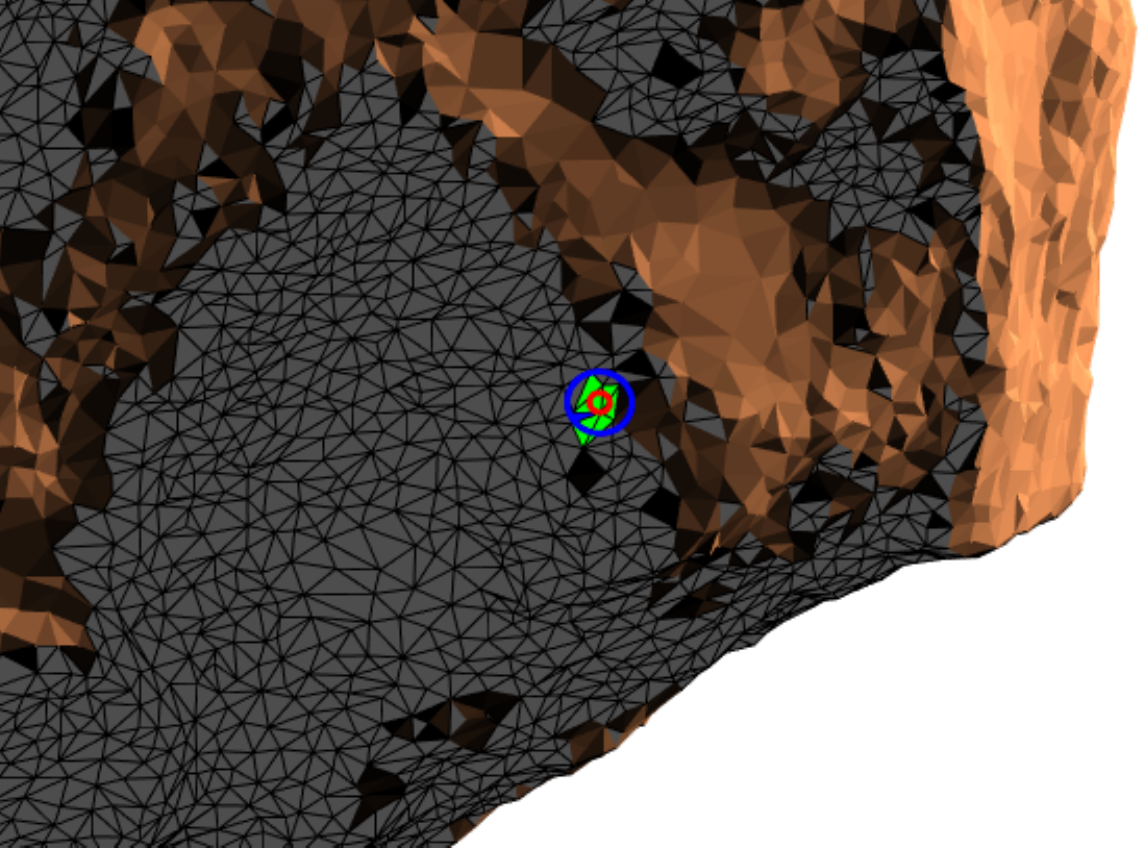}} & \scalebox{0.45}{\includegraphics{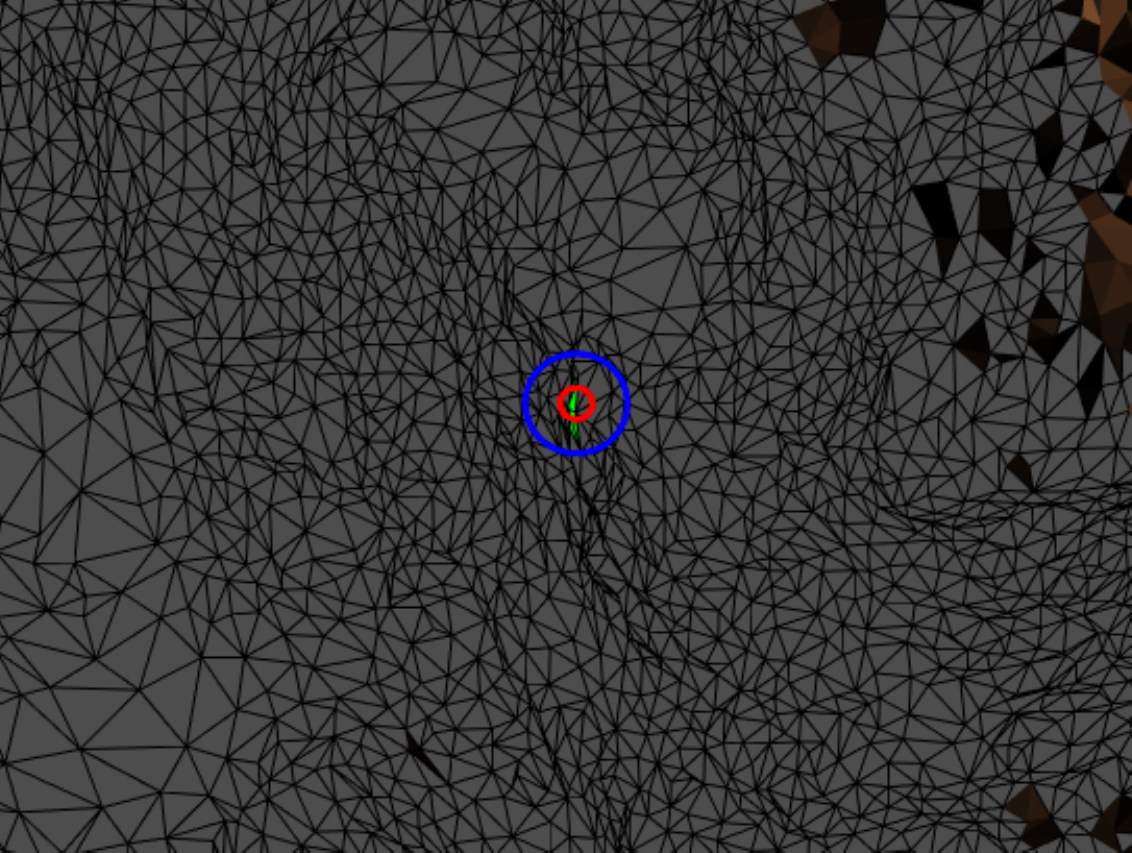}}\\
\end{tabular}
     \caption{Examples of Imhotep viewing conditions during acquisition of two different bins. The colour scale shows level of illumination, with grey areas being in darkness. 
The FWHM of the MIRO beams are shown as red (SMM) and blue (MM) circles. \emph{Left:}  Observations of a selected bin took place on 2014 December 15, at $23:47:09\,\mathrm{UTC}$, 
when \emph{Rosetta} was $20.4\,\mathrm{km}$ from the comet. The collapse area is viewed nearly face--on (the emergence angle is $e=13^{\circ}$).  \emph{Right:}  Observations of a rejected 
bin took place on 2015 January 20, at $01:27:37\,\mathrm{UTC}$, when \emph{Rosetta} was $27.7\,\mathrm{km}$ from the comet. Here, the collapse region is viewed edge--on and does not fill the beams.}
     \label{fig_bin_selection}
\end{figure*}

2014 December and 2015 January contributed 8 and 4 bins, respectively. The viewing geometries during each set of observations 
were scrutinised individually, in order to weed out those where the beams sampled too much adjacent terrain, sometimes with drastically different 
illumination conditions (e.~g., the collapse site being in shadow, but the beams including foreground or background fully illuminated terrain). 
Among the 2014 December bins, five had to be rejected, while three were selected. Figure~\ref{fig_bin_selection} (left panel) shows an example 
of the viewing conditions for a selected bin. In this case, the collapse site is clearly visible, it lies entirely in darkness, and fills most of the MIRO beams. Among the 2015 
January bins, three had to be rejected, while only one was selected. Figure~\ref{fig_bin_selection} (right panel) shows an example 
of the viewing conditions of a rejected bin. Here, the collapse site is viewed so obliquely that it only fills a small fraction of the beams. 

The final Imhotep temperature curve thus consisted of merely four bins, sampling early and late morning, as well as late night. 
The first observations used for the curve were acquired on 2014 December 8, and the last on 2015 January 15, thus spanning 
a period of 38 days. Note, that the two late--night bins near $d_{\rm n}=357.23$ were acquired at almost the same rotational 
phase but one week apart in real time. The bins are shown in Fig,~\ref{fig_imhotep_dust}. The MM bins (left panel) differ by 
$4.3\,\mathrm{K}$, i.e., their $\pm 2.5\,\mathrm{K}$ error bars overlap. The SMM bins (right panel) differ by merely $0.4\,\mathrm{K}$. 
The reason for the differences is partially due to the dependence of emitted radiance on emergence angle (being $e=14^{\circ}$ 
and $e=24^{\circ}$ for the two bins), and partially because of a small level of temperature dispersion within the beams (this is particularly 
the case for the larger MM beam). This exemplifies that the near--surface temperatures are similar from one nucleus revolution to the 
next, as assumed when the time--shift were made. A further example concerns the late--morning bin, observed on 2015 January 6, or 
about 29 nucleus revolutions after the initiation of the master period. It is the only one acquired in full daylight, and sensitive to the applied 
solar flux at the time (the other points where acquired when only nucleus self--heating contributed flux). The master--period illumination flux at this rotational phase was 
$74.3\,\mathrm{J\,m^{-2}\,s^{-1}}$, whereas the actual level of illumination at the time of observation was $70.3\,\mathrm{J\,m^{-2}\,s^{-1}}$. 
This is a difference of 5 per cent in flux, corresponding to a difference of $2.6\,\mathrm{K}$ in terms of equilibrium temperature. This 
illustrates that usage of master period fluxes at the time of observation introduces small errors.

\subsubsection{Refractories only} \label{sec_results_Imhotep_dust}

The brightness of the Imhotep collapse site, as well as its spectrally blue colour  \citep{augeretal15a,pommeroletal15}, excludes that the surface is 
covered by a dust mantle. Yet, it is interesting to compare its MIRO antenna temperatures with those predicted by \textsc{nimbus} and \textsc{themis} 
models for refractory material. The reason is primarily that the empirical antenna temperature curve consists of few bins, and it is important to understand 
whether this type of models can fit those particular available parts of the curve. If fits are possible, the evidence for presence of sublimating ice that MIRO might 
bring (in addition to the OSIRIS brightness and colour information), cannot be considered particularly strong. However, if dust--only models do not fit, but models 
with sublimating ice work better, it could be considered a MIRO--based verification that ice is exposed at the surface, that is independent from that of OSIRIS.

\begin{figure*}
\centering
\begin{tabular}{cc}
\scalebox{0.45}{\includegraphics{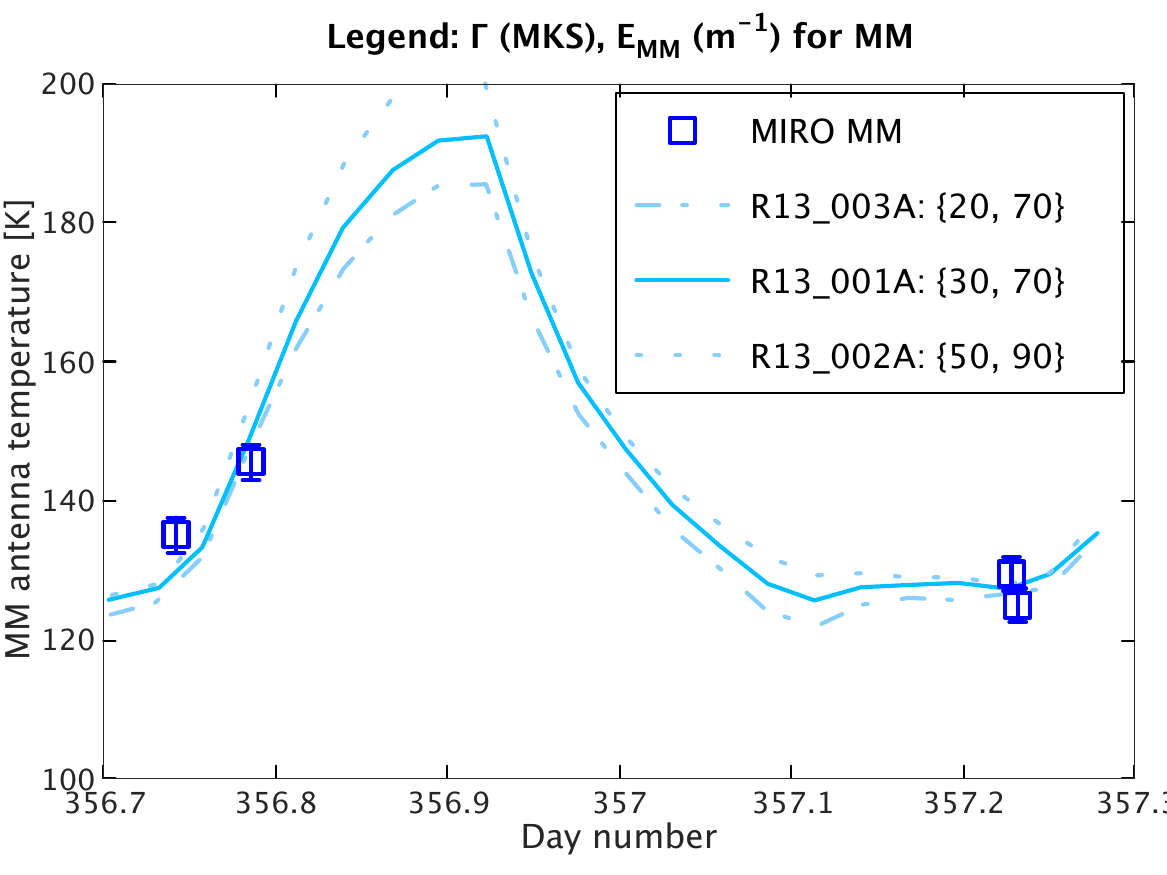}} & \scalebox{0.45}{\includegraphics{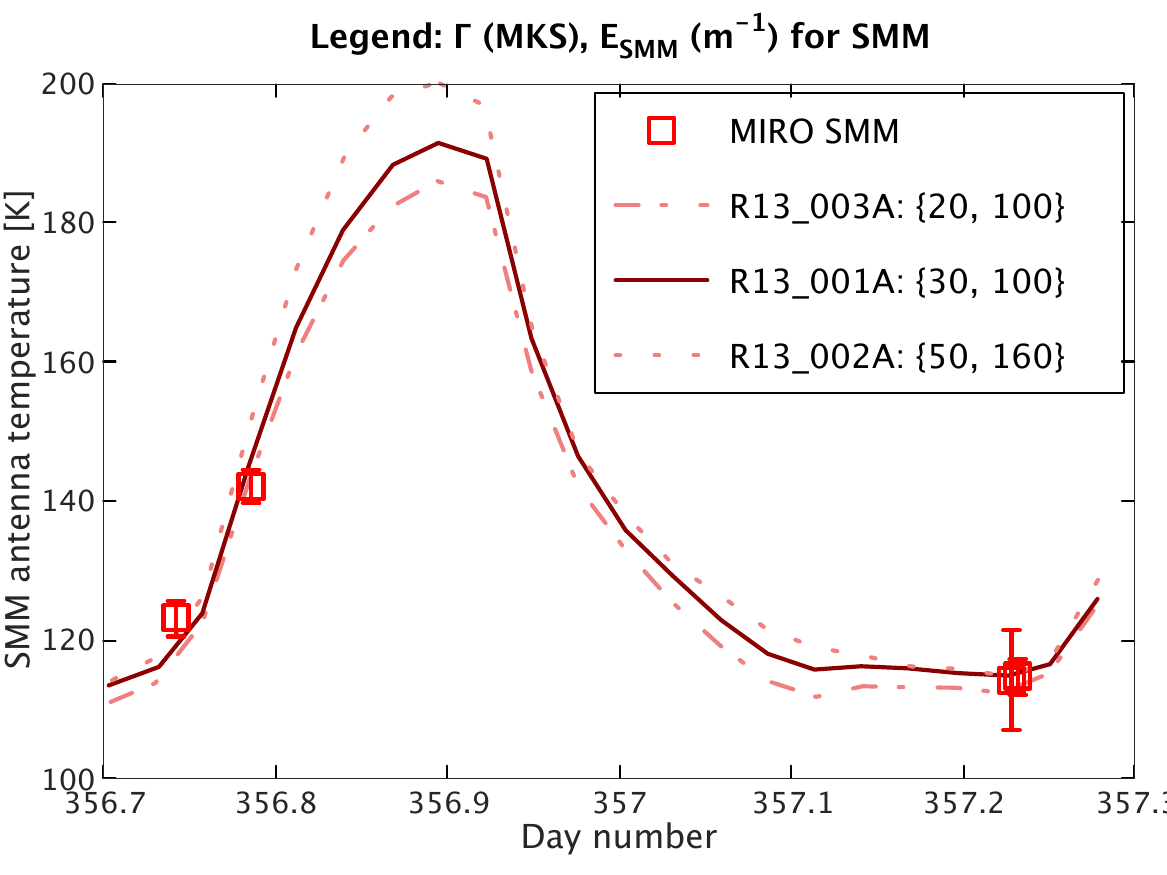}}\\
\end{tabular}
     \caption{Imhotep data compared with refractories--only models. Using different porosities $\psi=\{0.805,\,0.790,\,0.770\}$ here results in different bulk densities $\rho_{\rm bulk}=\{634,\,683,\,748\}\,\mathrm{kg\,m^{-3}}$, 
and Hertz factors $h=\{3.3\cdot 10^{-4},\,7.1\cdot 10^{-4},1.7\cdot 10^{-3}\}$. The resulting diurnal thermal inertia ranges are 13--$24\,\mathrm{MKS}$, 16--$35\,\mathrm{MKS}$, 
and 35--$53\,\mathrm{MKS}$ (referred to as 20, 30, and $50\,\mathrm{MKS}$ cases in the text). \emph{Left:} MM antenna temperatures. \emph{Right:} SMM antenna temperatures.}
     \label{fig_imhotep_dust}
\end{figure*}

Three \textsc{nimbus} models with different porosities ($0.770\leq\psi\leq 0.805$) where considered (see Fig.~\ref{fig_imhotep_dust} for additional parameter values). Because the \citet{shoshanyetal02} formalism was used to 
calculate the Hertz factor (the reduction of heat conductivity with respect to that of a compact substance due to porosity), this resulted in models having 
thermal inertia fluctuating around $\sim 20$, $\sim 30$, and $\sim 50\,\mathrm{MKS}$. This range is relevant based on previous work on the thermal 
inertia of 67P \citep[e.~g.][Paper~I]{schloerbetal15,marshalletal18,davidssonetal22b}. First, attempts were made to fit the MM data. $E_{\rm MM}$ was 
adjusted in \textsc{themis} until the model fitted the late--night bins. The result can be seen in Fig.~\ref{fig_imhotep_dust} (left panel). For the lowest thermal inertia the 
model is $2.3\,\mathrm{K}$ warmer than the nominal late--morning bin (at 66 per cent of the noon illumination flux of $145\,\mathrm{J\,m^{-2}\,s^{-1}}$), and barely consistent with its upper error margin. 
Increasing the thermal inertia only makes the model warmer (by $\leq 8.7\,\mathrm{K}$). However, all models are 4.3--$6.9\,\mathrm{K}$ too cold at the early morning bin (at 10 per cent of the noon illumination flux), the worst being the model with low thermal inertia. 
In essence, the modelled morning slope is too steep compared with the data, when $E_{\rm MM}$ is chosen to force a fit late at night. The best such model (R13\_001A with $\sim 30\,\mathrm{MKS}$) 
has $\chi_{\rm MM}^2=8.9$ for $E_{\rm MM}=70\,\mathrm{m^{-1}}$. This residual can be reduced somewhat if relaxing the requirement that the model curve should pass between the two late--night bins. 
Model R13\_003A with $\sim 20\,\mathrm{MKS}$ has $\chi_{\rm MM}^2=6.5$ for $E_{\rm MM}=50\,\mathrm{m^{-1}}$.

A similar investigation was made at the SMM wavelength (see Fig.~\ref{fig_imhotep_dust}, right panel), with similar results. When all models were made to fit late at night by adjusting $E_{\rm SMM}$, 
they were under--shooting in the early morning (by 2.5--$5.7\,\mathrm{K}$, improving as $\Gamma$ increased), and over--shooting in the late morning (by 1.3--$9.3\,\mathrm{K}$, 
deteriorating as $\Gamma$ increased). Again, the modelled morning slope appears to be steeper than the data and this problem cannot be fixed merely by $\Gamma$ and $E_{\rm MM}$ adjustments. 
The best of these models where R13\_001A with $\sim 30\,\mathrm{MKS}$, with $\chi_{\rm SMM}^2=4.9$ for $E_{\rm SMM}=100\,\mathrm{m^{-1}}$. By relaxing the requirement to fit the late--night bins, 
a somewhat lower $\chi_{\rm SMM}^2=3.1$ for $E_{\rm SMM}=80\,\mathrm{m^{-1}}$ could be obtained for the same thermophysical model. 

In terms of simultaneous MM and SMM fits, model R13\_003A is the best with $Q_{\rm MM}=0.011$ for $E_{\rm MM}=50\,\mathrm{m^{-1}}$ and $Q_{\rm SMM}=0.017$ for $E_{\rm SMM}=100\,\mathrm{m^{-1}}$, 
noting that a $Q>0.01$ success--criterion has been applied by \citet{davidssonetal22b} and in Paper~I based on recommendations by \citet{pressetal02}. Therefore, in a blind test, these MIRO observations 
would not be sufficient to exclude that a dust mantle is being observed. The fact that exposed ice fills most of the beams is therefore entirely based on OSIRIS observations.

\subsubsection{Refractories and $\mathit{H_2O}$} \label{sec_results_Imhotep_dust_H2O}

The OSIRIS observations suggest that the Imhotep collapse site has surface ice at the time of observation. Therefore, a number of \textsc{nimbus} models 
considered mixtures of refractories and water ice. First, a dust/water--ice mass ratio $\mu=1$ and diffusivity based on tube lengths and radii $\{L_{\rm p},\,r_{\rm p}\}=\{100,\,10\}\,\mathrm{\mu m}$ 
and tortuosity $\xi=1$ were applied (see Paper~I for their relation to $\mathcal{D}$), based on previous reproductions of the 67P water rate production curve \citep{davidssonetal22}. It was assumed that the original material had 
contained carbon dioxide at a molar abundance $\mathrm{CO_2/H_2O}=0.3$ relative to water \citep[][Paper~I]{davidssonetal22}, and had the same 
bulk density $\rho_{\rm bulk}=535\,\mathrm{kg\,m^{-3}}$ as the nucleus \citep{preuskeretal17}. Removal of the $\mathrm{CO_2}$, without any other structural changes 
than the resulting increase of porosity $\psi$, yielded $\psi=0.73$ and $\rho_{\rm bulk}=390\,\mathrm{kg\,m^{-3}}$ for the remaining dust/water mixture. 
The heat conductivity was corrected for porosity according to Hertz factors $h$ following \citet{shoshanyetal02}, resulting in a thermal inertia reaching $\sim 100\,\mathrm{MKS}$ at day. If water ice 
is removed from the surface layer, a $\psi=0.94$ porosity dust mantle is left behind. A lower limit $h\geq h_{\rm min}=h(\psi=0.79)$ was enforced to prevent the dust mantle from having 
a thermal inertia falling below $\sim 20\,\mathrm{MKS}$. 

\begin{figure*}
\centering
\begin{tabular}{cc}
\scalebox{0.45}{\includegraphics{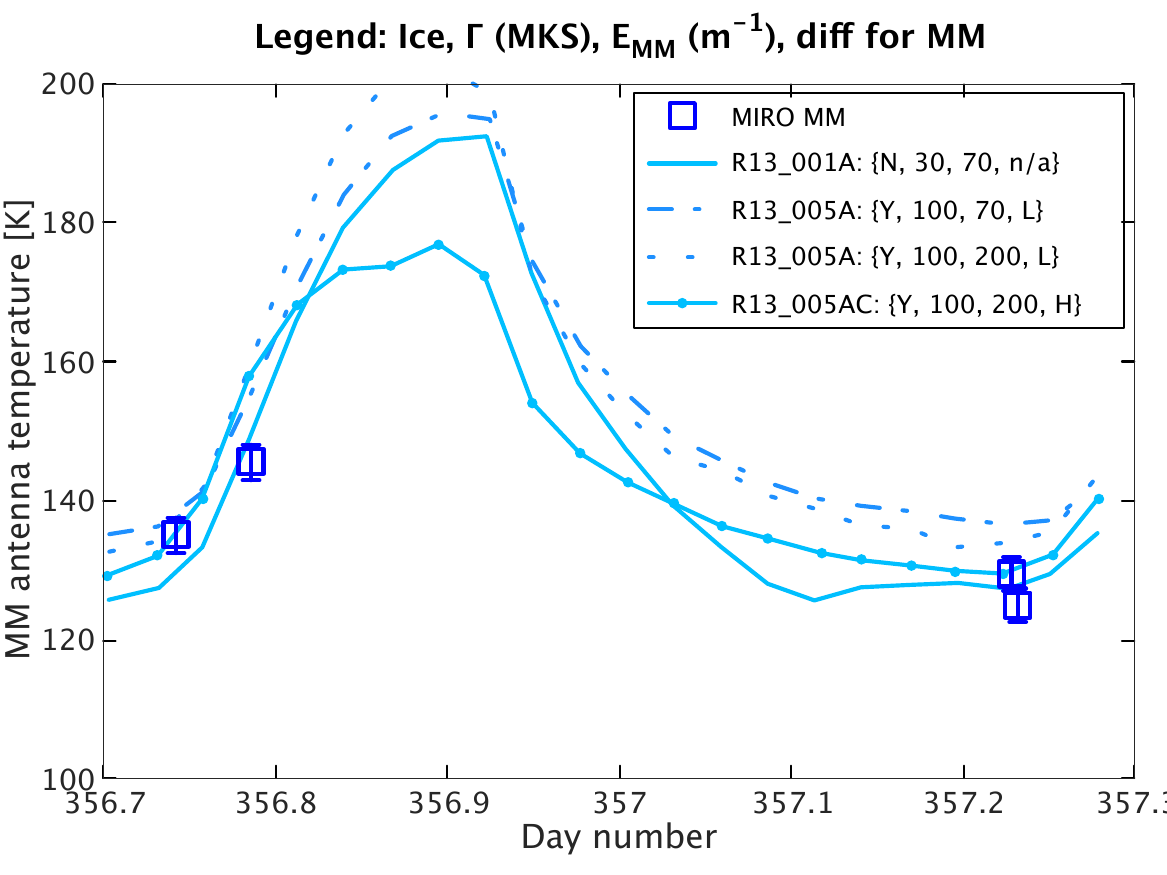}} & \scalebox{0.45}{\includegraphics{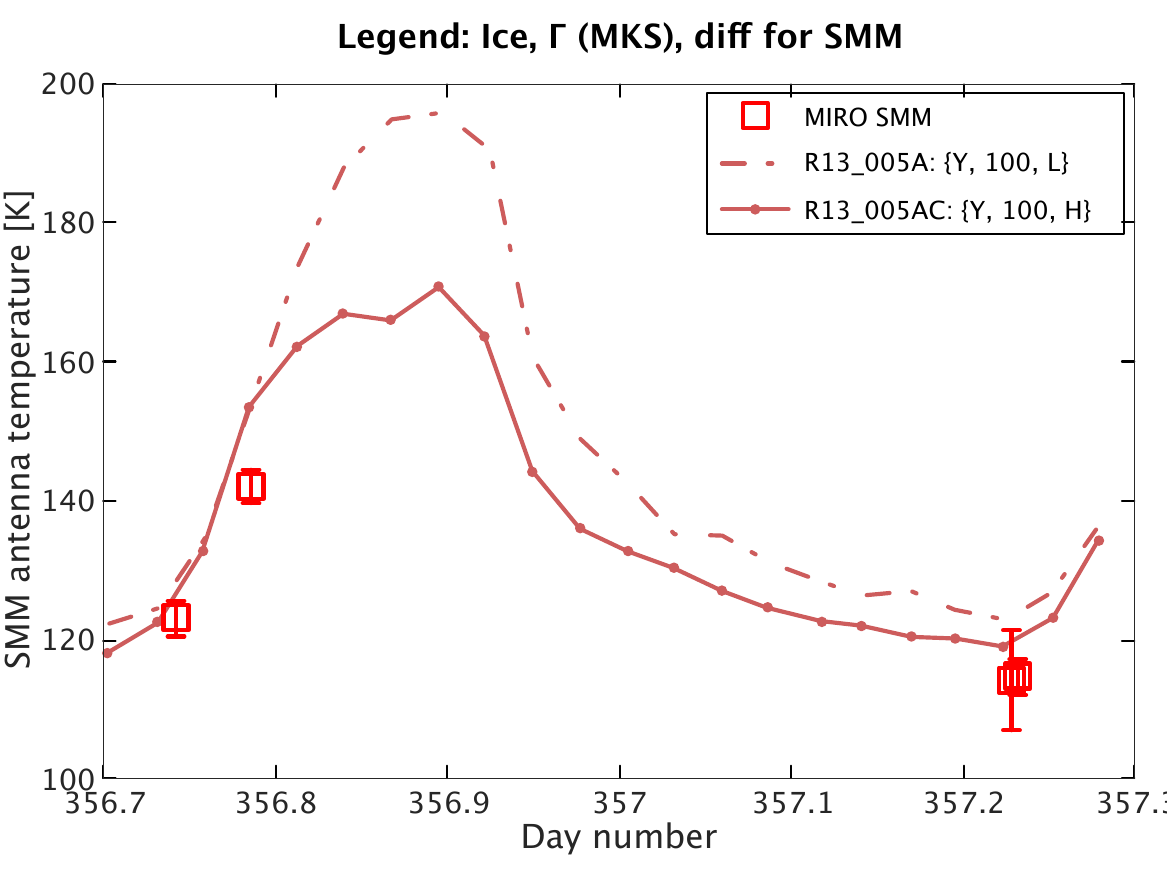}}\\
\end{tabular}
     \caption{Imhotep data compared with refractories$+\mathrm{H_2O}$ models. The legends indicate whether a given model considered 
water ice (Y/N for yes/no), the approximate thermal inertia $\Gamma$, the applied extinction coefficient ($E_{\rm MM}$ or $E_{\rm SMM}$), and 
whether the diffusivity (diff) was low (L) with $\{L_{\rm p},\,r_{\rm p}\}=\{100,\,10\}\,\mathrm{\mu m}$ and $\xi=1$, high (H) with $\{L_{\rm p},\,r_{\rm p}\}=\{10,\,1\}\,\mathrm{cm}$ and $\xi=1$, or 
not relevant (n/a for not applicable). \emph{Left:} MM data and model antenna temperatures. \emph{Right:}  SMM data and model antenna temperatures. Both models have $E_{\rm SMM}=600\,\mathrm{m^{-1}}$.}
     \label{fig_imhotep_dust_H2O}
\end{figure*}

Such a medium was modelled without erosion during a period of three weeks just prior to the master period, to understand the level of water activity. In this time, the water withdrew 
merely $\sim 2.5\,\mathrm{mm}$, showing that the water ice was almost dormant. The main effect of adding water is therefore to produce a denser medium with higher volumetric 
heat capacity, and lower porosity which leads to a higher heat conductivity, thus a higher thermal inertia. Figure~\ref{fig_imhotep_dust_H2O} (left panel) compares the MM curves for R13\_001A 
without water ice (from Fig.~\ref{fig_imhotep_dust}) with R13\_005A (considering water ice, including at the surface) for the same extinction coefficient $E_{\rm MM}=70\,\mathrm{m^{-1}}$. Increasing the thermal inertia has the effect of 
elevating the MM antenna temperature, and more significantly at night than at day. The quasi--fit of the dust model is thereby ruined ($\chi_{\rm MM}^2$ increases from 8.9 to 47). 
Increasing the extinction coefficient has the effect of lowering nighttime temperatures, but elevate daytime temperatures. Figure~\ref{fig_imhotep_dust_H2O} (left panel) shows 
that the error bars of the late night bins are barely reached when $E_{\rm MM}=200\,\mathrm{m^{-1}}$, and additional simulations show that the antenna temperature is not reduced further by stronger extinction. 
However, because of increased discrepancies in the late morning, this attempt to improve the fit late at night only leads to a somewhat higher $\chi_{\rm MM}^2=51$. An $E_{\rm MM}$ this high 
may be nonphysical, because it would suggest an opacity so high that the MM channel only probes the top $\sim 5\,\mathrm{mm}$. Furthermore, \citet{schloerbetal15} found an average 
$E_{\rm MM}=26^{+11}_{-6}\,\mathrm{m^{-1}}$, \citet{davidssonetal22b} fitted $E_{\rm MM}=25$--$80\,\mathrm{m^{-1}}$ for a location in Hapi, and Paper~I had 
$E_{\rm MM}=20$--$70\,\mathrm{m^{-1}}$ for the Aswan collapse site. In comparison, $E_{\rm MM}=200\,\mathrm{m^{-1}}$ is indeed rater high. 

Figure~\ref{fig_imhotep_dust_H2O} (left panel) shows that R13\_005A has similar problems at SMM wavelengths, with the antenna temperature $T_{\rm A}$ being too high at all bins for 
$E_{\rm SMM}=600\,\mathrm{m^{-1}}$ (and $T_{\rm A,SMM}$ would not decrease more if $E_{\rm SMM}$ was further increased). For this model, $\chi_{\rm SMM}^2=40$.

It therefore seems that a dust/water mixture performs worse than dust alone, at least when the diffusivity is as low as suggested by $\{L_{\rm p},\,r_{\rm p}\}=\{100,\,10\}\,\mathrm{\mu m}$ 
and $\xi=1$. This is surprising, considering that the OSIRIS observations suggests that water ice indeed is present at the surface at the collapse site. Therefore, higher diffusivities 
were considered to increase the degree of sublimation cooling, including one based on $\{L_{\rm p},\,r_{\rm p}\}=\{10,\,1\}\,\mathrm{cm}$ and $\xi=1$. Such large values appear to have been common post--perihelion on the northern hemisphere 
\citep{davidssonetal22}, presumably due to the replenishment of airfall debris around perihelion, that have typical sizes in the millimetre--decimetre range 
\citep{mottolaetal15,pajolaetal16,pajolaetal17b}. On the inbound orbit, the much lower diffusivity used above as default is typical, suggesting some sort of pulverisation and settling of material 
around aphelion. If such a transition indeed took place,  some locations may have persisted at high diffusivities for longer. For example, \citet{davidssonetal22b} found that a location in 
Hapi appears to have had a three orders of magnitude drop in diffusivity between 2014 October and November. However, airfall debris would not accumulate on a steep cliff side. 
Still, there might be other mechanisms having similar effects. In Paper~I it was found that $\{L_{\rm p},\,r_{\rm p}\}=\{1,\,0.1\}\,\mathrm{cm}$ and $\xi=1$ characterised the Aswan collapse 
site five months after the wall fell, and I speculated that vigorously sublimating $\mathrm{CO_2}$ (estimated to be located $0.4\pm 0.2\,\mathrm{cm}$ below the surface) was responsible for 
`puffing up' the surface layer and increase the diffusivity. Here, it is assumed that a similar mechanism might have been active at the Imhotep site. 

Increasing the diffusivity to a level determined by $\{L_{\rm p},\,r_{\rm p}\}=\{10,\,1\}\,\mathrm{cm}$ and $\xi=1$ intensifies sublimation only slightly. In three weeks, without erosion, the 
dust mantle thickness increases from $\sim 0.25$ to $\sim 1.3\,\mathrm{cm}$. Adding $Q_{\rm e}=2Q_{\rm H_2O}\,\mathrm{(kg\,m^{-2}\,s^{-1})}$ erosion during the last week before 
the master period brings the water ice to the surface and maximises the cooling due to its sublimation. Figure~\ref{fig_imhotep_dust_H2O} (left panel) shows that the MM antenna temperature becomes 
somewhat lower (R13\_005AC), particularly during day. The late night MM antenna temperature approaches the data bins only if $E_{\rm MM}\geq 100\,\mathrm{m^{-1}}$ (the $200\,\mathrm{m^{-1}}$ 
case is shown here to allow direct comparison with the low--diffusivity case R13\_005A). Yet, the model is $\sim 13\,\mathrm{K}$ warmer than the late--morning bin, which is typical of all 
models including water. At SMM wavelengths ( Fig.~\ref{fig_imhotep_dust_H2O}, right panel) there are similar discrepancies, reaching $\sim 12\,\mathrm{K}$. Both displayed R13\_005AC 
models have $\chi_{\rm MM}^2\approx\chi_{\rm SMM}^2\approx 30$. Lowering to $E_{\rm MM}=50\,\mathrm{m^{-1}}$ and $E_{\rm SMM}=200\,\mathrm{m^{-1}}$ worsen the fits late at 
night (6--$9\,\mathrm{K}$ discrepancies) but improve them in the late morning, lowering overall residuals to $\chi_{\rm MM}^2\approx \chi_{\rm SMM}^2\approx 18$. Therefore, not even 
the strongest achievable water sublimation cooling is capable of cancelling the effect of increased thermal inertia, introduced by the presence of water itself. However, it indicates that 
sublimation cooling by the more volatile $\mathrm{CO_2}$ is worth testing.

\begin{figure*}
\centering
\begin{tabular}{cc}
\scalebox{0.45}{\includegraphics{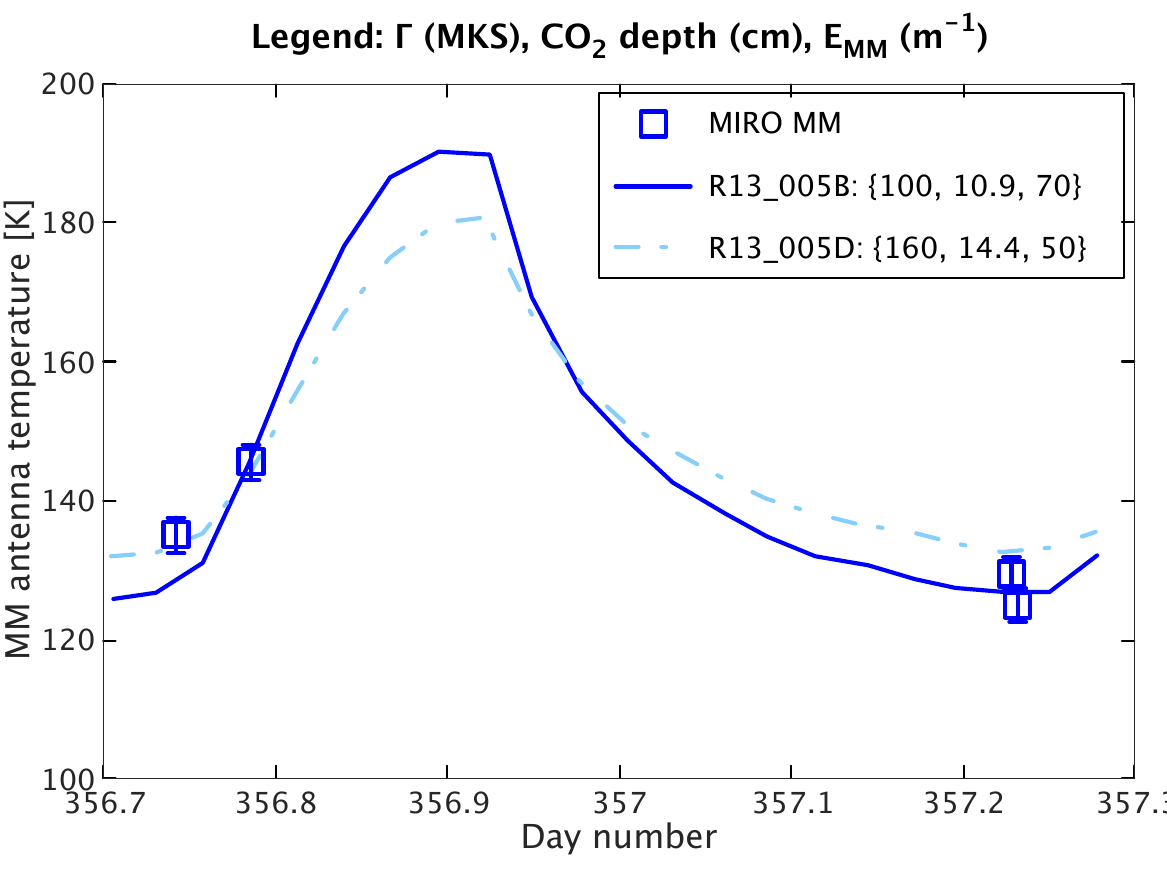}} & \scalebox{0.45}{\includegraphics{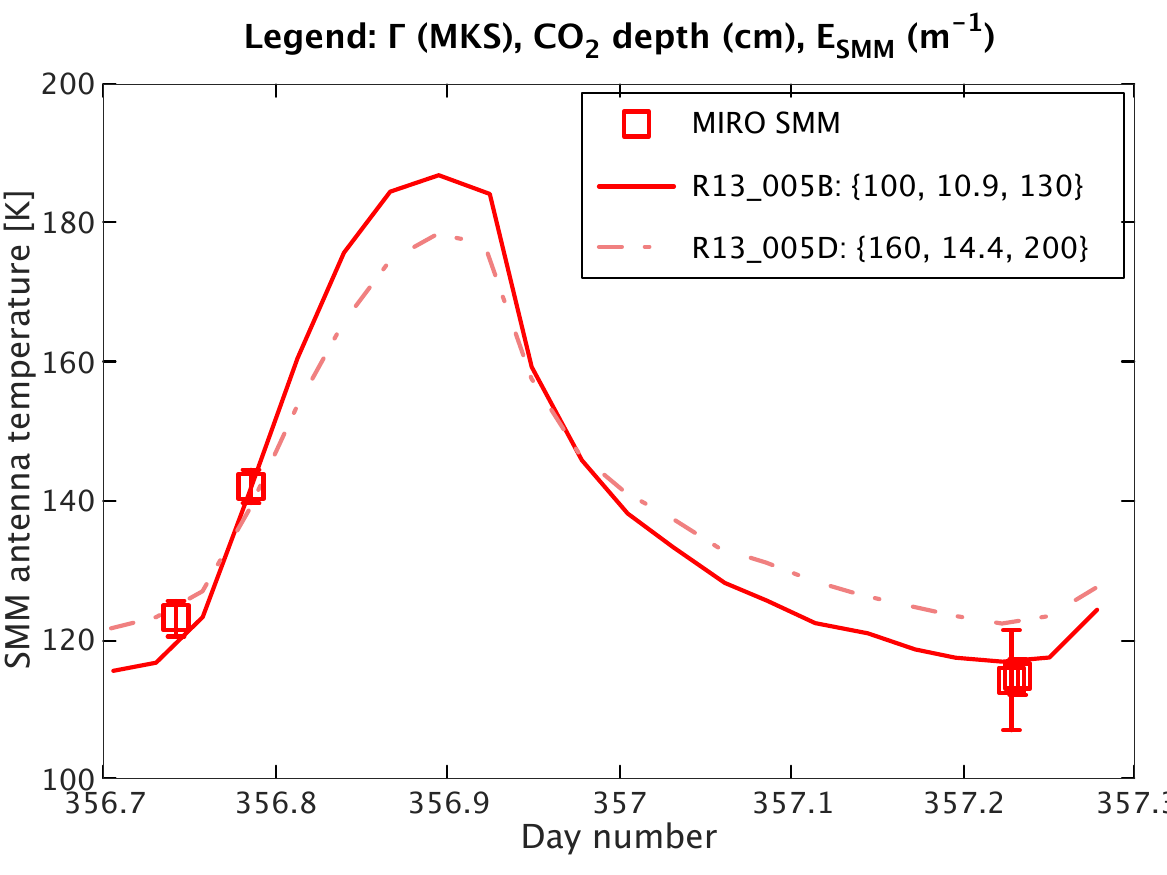}}\\
\end{tabular}
     \caption{Imhotep data compared with refractories$+\mathrm{H_2O}+\mathrm{CO_2}$ models. Both models have $\mu=1$, $\rho_{\rm bulk}=535\,\mathrm{kg\,m^{-3}}$ below the $\mathrm{CO_2}$ sublimation 
front,  $\{L_{\rm p},\,r_{\rm p}\}=\{10,\,1\}\,\mathrm{cm}$ and $\xi=1$. Model R13\_005B has $\mathrm{CO_2/H_2O}=0.30$ and model R13\_005D has $\mathrm{CO_2/H_2O}=0.15$, resulting in different 
thermal inertia as indicated by the legends. Furthermore, different erosion rates are applied, which places the $\mathrm{CO_2}$ front at different depths. \emph{Left:} MM data and simulations. \emph{Right:} SMM data and simulations.}
     \label{fig_imhotep_dust_H2O_CO2}
\end{figure*}

\subsubsection{Refractories, $\mathit{H_2O}$, and $\mathit{CO_2}$} \label{sec_results_Imhotep_dust_H2O_CO2}

The first model that included $\mathrm{CO_2}$ (R13\_005B) had the default parameters $\mu=1$, molar $\mathrm{CO_2/H_2O}=0.3$, $\rho_{\rm bulk}=535\,\mathrm{kg\,m^{-3}}$, 
$\{L_{\rm p},\,r_{\rm p}\}=\{100,\,10\}\,\mathrm{\mu m}$ and $\xi=1$. Such a medium was first propagated during a period of three weeks from an initially homogeneous state, assuming 
that no erosion took place. At the time of the master period, the $\mathrm{CO_2}$ sublimation front had withdrawn to a depth of $10.9\,\mathrm{cm}$, illustrating its higher 
sublimation rate with respect to water. Interestingly, no dust mantle formed in this time period. The reason is the cooling effect of $\mathrm{CO_2}$, that quenched the water 
production rate substantially with respect to models R13\_001A and R13\_001AC.

The $\mathrm{CO_2}$--free top layer, consisting of refractories and water ice, has the same $\sim 100\,\mathrm{MKS}$ thermal inertia as the models in section~\ref{sec_results_Imhotep_dust_H2O}. 
Yet, because of the stronger cooling of $\mathrm{CO_2}$ sublimation compared to that of $\mathrm{H_2O}$, the temperature in the near--surface layer is reduced. Thereby, the 
antenna temperatures are more easy to reproduce, than for media where the only volatile is water ice. That can be seen in the left panel of Fig.~\ref{fig_imhotep_dust_H2O_CO2}, where 
the model has $\chi_{\rm MM}^2=8.9$ and $Q_{\rm MM}=0.003$ when forced to pass between the two late--night bins by setting $E_{\rm MM}=70\,\mathrm{m^{-1}}$. The model works even more convincingly 
at SMM, seen in the right panel of Fig.~\ref{fig_imhotep_dust_H2O_CO2}. Here, the residuals are characterised by $\chi_{\rm SMM}^2=3.2$ and $Q_{\rm SMM}=0.072$ when $E_{\rm SMM}=130\,\mathrm{m^{-1}}$. 

A model (R13\_005C) with $Q_{\rm e}=1.5(Q_{\rm H_2O}+Q_{\rm CO_2})$ erosion initiated a week before the master period was tested to see if a more 
shallow $\mathrm{CO_2}$ front (at $6.9\,\mathrm{cm}$) was compatible with the data. In this case, the model curve was too cold, with $\chi_{\rm MM}^2=31$ 
and $\chi^2=18$ for $E_{\rm MM}=50\,\mathrm{m^{-1}}$ and $E_{\rm SMM}=80\,\mathrm{m^{-1}}$ respectively. Another model (R13\_005D) tested the effect of 
increasing the thermal inertia in the dust/water layer to $\Gamma=160\pm 10\,\mathrm{MKS}$ by lowering the $\mathrm{CO_2}$ abundance to $\mathrm{CO_2/H_2O=0.15}$ 
while maintaining $\rho_{\rm bulk}=535\,\mathrm{kg\,m^{-3}}$. With $Q_{\rm e}=3.0(Q_{\rm H_2O}+Q_{\rm CO_2})$ erosion, this model had the $\mathrm{CO_2}$ at 
a depth of $14.4\,\mathrm{cm}$ during the master period. This combination of higher thermal inertia and weaker $\mathrm{CO_2}$ sublimation cooling had the 
interesting effect of reducing the morning curve slope, thereby bringing it closer to that of the data (see Fig.~\ref{fig_imhotep_dust_H2O_CO2}). However, such a model 
starts to lose contact with the late--night bins, yielding $\chi_{\rm MM}^2=12$, $\chi_{\rm SMM}^2=13$ for $E_{\rm MM}=50\,\mathrm{m^{-1}}$ and $E_{\rm SMM}=200\,\mathrm{m^{-1}}$, 
respectively. 

If the residuals for the various Imhotep models are considered (see Table~\ref{tab1} for a summary) it appears that $\mathrm{CO_2}$--rich media perform nearly as 
well as the dust--only media. The cooling provided by $\mathrm{CO_2}$ sublimation compensates for the higher thermal inertia introduced by adding $\mathrm{H_2O}$ and $\mathrm{CO_2}$ 
ices. It is also evident that the less volatile water ice is not capable of such compensation on its own. Unfortunately, the placement of the few available bins prevents us from 
clearly distinguishing between the dust--only and dust/$\mathrm{H_2O}$/$\mathrm{CO_2}$ cases. However, the additional observations by OSIRIS favours the exposure of ices at the 
surface at the Imhotep collapse site. Based on the available simulations, a tentative depth of the $\mathrm{CO_2}$ sublimation front is $11\pm 4\,\mathrm{cm}$.

\begin{table*}
\begin{center}
\begin{tabular}{||l|r|l|r|r|r|r|r|r||}
\hline
\hline
Model & $\mathrm{H_2O}$ & $\mathrm{CO_2}$ & $\Gamma\,\mathrm{(MKS)}$ & $\{L_{\rm p},\,r_{\rm p}\}$ & $E_{\rm MM}\,\mathrm{(m^{-1})}$ & $\chi_{\rm MM}^2$ & $E_{\rm SMM}\,\mathrm{(m^{-1})}$ &  $\chi_{\rm SMM}^2$\\
\hline
R13\_003A & N & N & 20 & -- & 50 & 6.5 & 100 & 5.7\\
R13\_001A & N & N & 30 & -- & 70 & 8.9 & 80 & 3.1\\
\hline
R13\_005A & Y & N & 100 & $\{100,\,10\}\,\mathrm{\mu m}$ & 100 & 44 & 600 & 40\\
R13\_005AC & Y & N & 100 & $\{10,\,1\}\,\mathrm{cm}$ & 50 & 18 & 200 & 17\\
\hline
R13\_005B & Y & Y ($10.9\,\mathrm{cm}$) & 100 & $\{100,\,10\}\,\mathrm{\mu m}$ & 70 & 8.5 & 130 & 3.2\\
R13\_005D & Y & Y ($14.4\,\mathrm{cm}$) & 160 & $\{100,\,10\}\,\mathrm{\mu m}$ & 50 & 12 & 200 & 13\\
\hline 
\hline
\end{tabular}
\caption{Summary of the \textsc{nimbus} and \textsc{themis} models for the Imhotep case that performed best for the three different types of composition (dust--only, dust$+\mathrm{H_2O}$, and 
dust$+\mathrm{H_2O}+\mathrm{CO_2}$). Y/N for yes/no on presence of the volatile in question. Note that the thermal inertia values applies to either dust or the dust$+\mathrm{H_2O}$ layer, depending on what is relevant (also for models that include $\mathrm{CO_2}$).}
\label{tab1}
\end{center}
\end{table*}

\subsection{The Hathor collapse site: 2014 Nov and Dec} \label{sec_results_Hathor}

The Hathor collapse site constitutes a larger target than that at Imhotep, measuring $\sim 105\,\mathrm{m}$ across. The highest concentration of 
high--resolution observations took place in 2014 November and December. First, the data acquired during the two months were considered separately. 
Whereas the 2014 November data set covers the night and forenoon parts of the diurnal curve, the 2014 December data set covers late afternoon, night, and 
morning. Because the antenna temperatures are similar at the portions of the curve where there is overlap, the two sets were merged to obtain a 
more complete diurnal curve. The combined dataset consisted of 171 individual 1--second continuum observations, acquired between 
November 8 and December 29 during a 50.6 day period. The observations took place when \emph{Rosetta} was 20.6--$30.4\,\mathrm{km}$ from the comet. 
A master period starting on $d_{\rm n}=337.0452$ was selected (2014 December 3, 01:05:05~UTC) and usage of $2.4\,\mathrm{min}$--wide binning intervals resulted 
in 12 data bins. Viewing and illumination conditions were scrutinised individually and all bins represented unobscured views of Hathor. For bins \#1--2 Hathor was fully 
illuminated, while for bins \#3--8 the region was in full darkness, considering the FWHM extensions of both the MM and SMM beams. The FWHM extension of the 
smaller SMM beam contained only dark terrain for bins \#9--10, and only faintly illuminated terrain for bins \#11--12. Because the SMM footprint did not include drastically 
different illumination conditions at any point, all 12 bins were selected for the SMM analysis. However, the larger MM footprint sampled some faintly illuminated 
terrain when most of the Hathor collapse site was in darkness (bins \#9--10), and it sampled some shadowed terrain when most of the collapse site was faintly illuminated 
(bins \#11--12). Therefore, bins \#9--12 were rejected in the MM analysis. 

Figure~\ref{fig_hathor_dust} shows the MM and SMM data in the left and right panels, respectively. To illustrate the consistency in antenna temperatures measured on different 
occasions but at similar nucleus rotational phases, it is noted that: 1) the three bins at $d_{\rm n}=337.31$--$337.35$ were acquired on November 10, December 18 and 29, yet 
agree to within $1.4\,\mathrm{K}$ (MM) and $2.2\,\mathrm{K}$ (SMM); 2) the three bins at $d_{\rm n}=337.42$--$337.45$ were acquired on November 8 and 29, and December 26, yet 
agree to within $0.5\,\mathrm{K}$ (MM) and $2.6\,\mathrm{K}$ (SMM). The four rejected MM bins at $d_{\rm n}=337.51$--$337.52$ are shown for completeness (they agree 
to within $2.3\,\mathrm{K}$). Interestingly, the daytime MM antenna temperatures are not significantly elevated above the nighttime data, i.~e., the MM curve seems to have a very 
small amplitude ($4.4\,\mathrm{K}$) compared to the SMM curve ($15.1\,\mathrm{K}$).

It is interesting to note that the Hathor MM antenna temperatures ($\sim 180\,\mathrm{K}$) are much hotter than those of Imhotep (130--$150\,\mathrm{K}$). 
This suggests that the physical conditions that are governing each site ought to be rather different.

\subsubsection{Refractories only} \label{sec_results_hathor_dust}

Though several bright and bluish $\leq 10\,\mathrm{m}$ spots are seen at the Hathor alcove \citep{thomasetal15a}, suggesting the presence of exposed water ice, 
most of the area has a dark dust coverage. The first step is therefore to test whether the MIRO antenna temperature curves can be reproduced by assuming an 
ice--free medium consisting of a porous assembly of refractory grains. Nominally, the \textsc{nimbus} simulations considered the standard temperature--dependent 
expressions for specific heat capacity and heat conductivity for compacted material used by \citet{davidsson21}. Assuming a compacted density $\rho_1=3250\,\mathrm{kg\,m^{-3}}$ 
for the dust monomer grains, and applying the \citet{shoshanyetal02} heat conductivity correction $h$ due to porosity, the medium was assigned a porosity $\psi=0.79$ in order to 
achieve a baseline thermal inertia of $\Gamma \approx 30\,\mathrm{MKS}$. The resulting bulk density of the porous dust mantle was $\rho_{\rm bulk}=682\,\mathrm{kg\,m^{-3}}$. 

This model (R14\_001A) was propagated from the 2012 May 23 aphelion of Comet 67P, up to and including the alcove master period, using the illumination conditions 
specific to this location in Hathor. This model had $\Gamma=33\pm 3\,\mathrm{MKS}$ during the master period (with the dispersion in thermal inertia caused by diurnal temperature variations). 
Other models considered 2 or 4 times lower or higher thermal inertia by multiplying or dividing $h$ by factors 4 or 16. These models used the R14\_001A conditions $\sim 28$ days prior to 
the master period as initial conditions before being propagated just beyond that period. Using \textsc{themis} to calculate the corresponding MM antenna temperature, the extinction 
coefficient $E_{\rm MM}$ was adjusted until the models roughly reproduced the nighttime data \citep[assuming a single--scattering albedo $w_{\rm MM}=0$, as expected for rocky 
material at these wavelengths;][]{garykeihm78}.

\begin{figure*}
\centering
\begin{tabular}{cc}
\scalebox{0.45}{\includegraphics{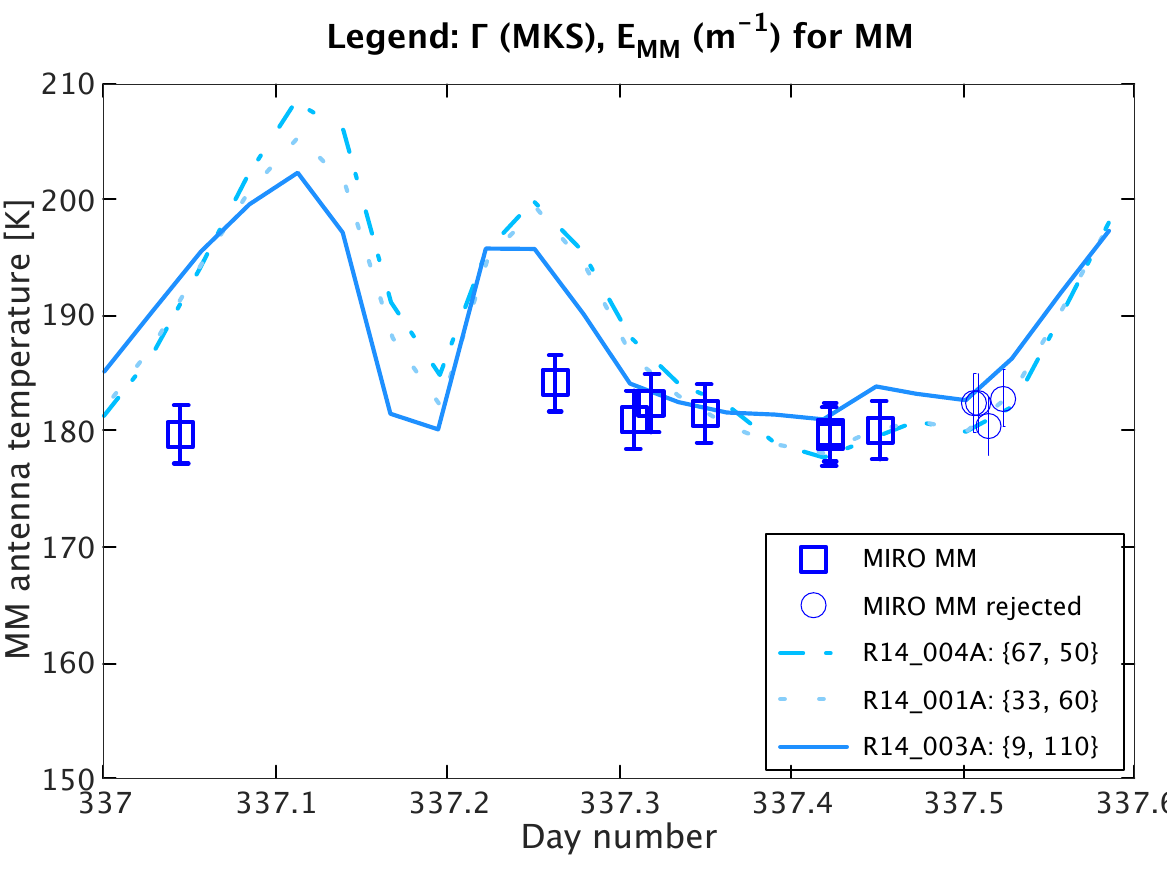}} & \scalebox{0.45}{\includegraphics{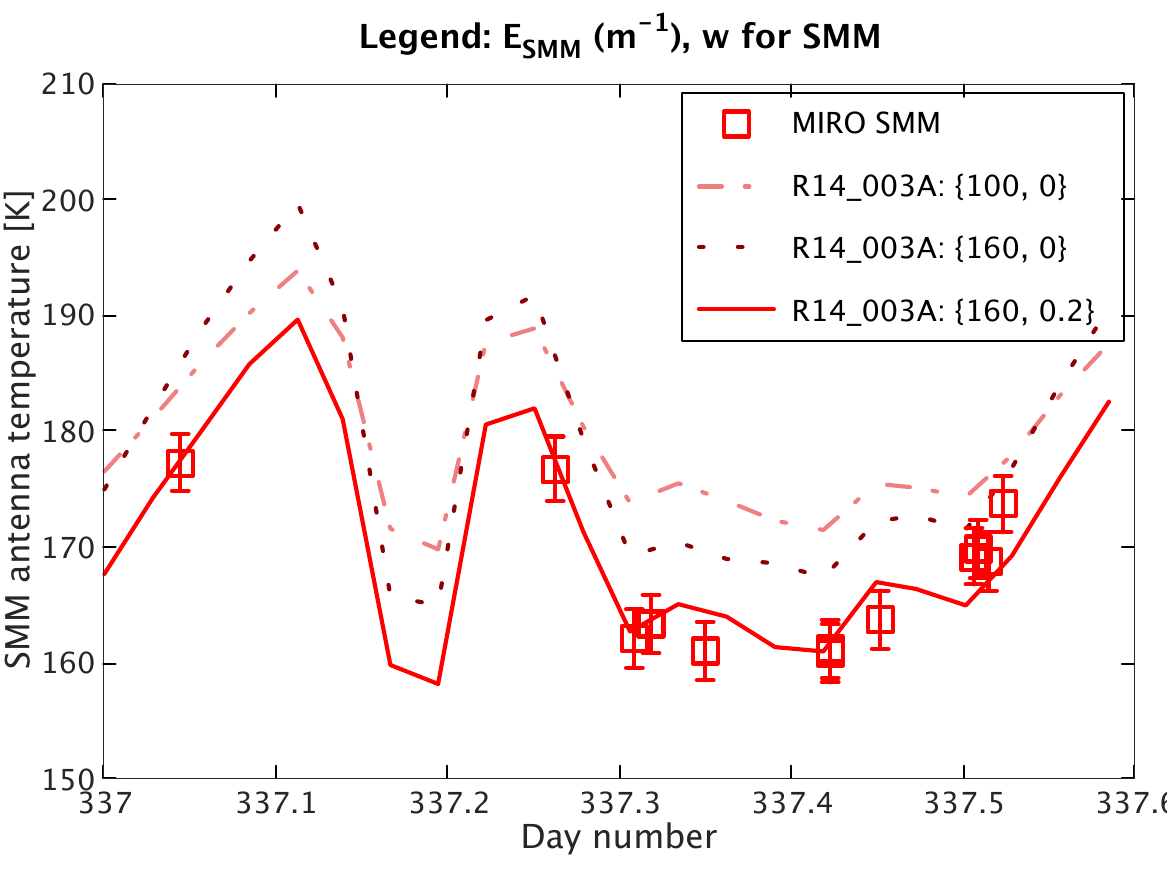}}\\
\end{tabular}
     \caption{Hathor data compared with refractories--only models. \emph{Left:} Measured and modelled MM antenna temperatures. Here, the effect of lowering the thermal inertia is shown, while adjusting 
$E_{\rm MM}$ until the nighttime data are reproduced. The slight reduction of the curve amplitude associated with thermal inertia reduction is not sufficient to reproduce the daytime temperatures even 
for the very low $\Gamma=9\pm 1\,\mathrm{MKS}$.}
     \label{fig_hathor_dust}
\end{figure*}

The results for a few models can be seen in Fig.~\ref{fig_hathor_dust} (left panel). Model R14\_004A with $\Gamma=63\pm 4\,\mathrm{MKS}$ roughly fits the nighttime data 
for $E_{\rm MM}=50\,\mathrm{m^{-1}}$, but the daytime temperatures are far too high. Note that the `double--peaked' antenna temperature curve is due to a shadowing 
episode during daytime, when the large lobe temporarily is blocking the Sun. By reducing the thermal inertia, the near--surface layer with substantial temperature variations (i.~e., the thermal skin depth) 
is shrinking. If this is done while keeping $E_{\rm MM}$ fixed, the layer probed by MIRO becomes increasingly isothermal, and the MM curve amplitude is thus reduced as $\Gamma$ is falling. 
However, the curve then approaches an average level that is too warm compared 
to the nighttime data. It can be brought down by increasing the extinction coefficient to $E_{\rm MM}=60\,\mathrm{m^{-1}}$ for  R14\_001A (with $\Gamma=33\pm 3\,\mathrm{MKS}$) and 
to $E_{\rm MM}=110\,\mathrm{m^{-1}}$ for R14\_003A (with $\Gamma=9\pm 1\,\mathrm{MKS}$). But then the measured radiance again primarily comes from within the skin depth. As the 
left panel of Fig.~\ref{fig_hathor_dust} shows, it is difficult to obtain a net amplitude reduction just by adjusting the thermal inertia. 

Adjusting the assumed bulk density $\rho_{\rm bulk}=682\,\mathrm{kg\,m^{-3}}$ does not help. As explained by \citet{schloerbetal15}, the same antenna temperature 
curve is obtained (at a given thermal inertia, and for homogeneous media) as long as the $\rho c/E$ ratio is fixed. To verify this, model R14\_001A with $\rho=682\,\mathrm{kg\,m^{-3}}$ and $E_{\rm MM}=50\,\mathrm{m^{-1}}$ 
was compared to a model R14\_005A with $\rho=169\,\mathrm{kg\,m^{-3}}$ and $E_{\rm MM}=10\,\mathrm{m^{-1}}$ (both having $\sim 30\,\mathrm{MKS}$). Having comparable $\rho c/E$ ratios, 
their MM antenna temperature curves were indeed very similar. If a bulk density $\rho$ (thus, heat capacity $\rho c$) is randomly chosen, and a fit to measured data cannot be obtained for any $E$, there is 
no point in trying other bulk density values because the achievable antenna temperature curves will be the same (except that a given curve will be obtained for a somewhat different $E$--value 
than previously). For this reason, a refractories--only model is not capable of reproducing the low--amplitude MM data.

At SMM wavelengths, the modelled antenna temperature is typically too high throughout the rotation period when $w_{\rm SMM}=0$, as illustrated for model R14\_003A in the 
right panel of Fig.~\ref{fig_hathor_dust}. In order to lower the nighttime modelled values towards the data, it is beneficial to increase the extinction coefficient. This is illustrated in the 
figure by the difference in applying $E_{\rm SMM}=100$ and $160\,\mathrm{m^{-1}}$. However, a large $E_{\rm SMM}$ also enhances the curve amplitude, making the daytime temperatures 
increasingly too high. The only way to achieve a match is to increase $E_{\rm SMM}$ to the point that the model curve has the correct amplitude (though at too high absolute temperatures), 
and then introduce scattering in the material by considering $w_{\rm SMM}>0$ (this pulls down the curve rather evenly at all rotational phases). Among the various models considered (50 combinations of 
$\Gamma$, $E_{\rm SMM}$, and $w_{\rm SMM}$), the best fit ($Q_{\rm SMM}=0.11$) was obtained for $\Gamma=9\pm 1\,\mathrm{MKS}$, $E_{\rm SMM}=160\,\mathrm{m^{-1}}$, and $w_{\rm SMM}=0.2$. 
Acceptable solutions ($Q_{\rm SMM}>0.01$) included the entire $120\leq E_{\rm SMM}\leq 180\,\mathrm{m^{-1}}$ range for  $\Gamma=9\pm 1\,\mathrm{MKS}$ and $w_{\rm SMM}=0.2$, as well 
as marginal fits ($Q_{\rm SMM}\approx 0.04$) for $\Gamma=16\pm 2\,\mathrm{MKS}$, $E_{\rm SMM}=110\pm 10\,\mathrm{m^{-1}}$, and $w_{\rm SMM}=0.2$.

Despite these SMM solutions, refractory--only models are here considered failed. The most important reason is that there are no MM solutions. But additionally, the high SMM modelled antenna 
temperatures (here dealt with by introducing significant scattering), could be an artificial consequence of having neglected sublimation cooling. To test whether cooling by subliming ices 
might produce more compelling simultaneous MM and SMM solutions, the following section therefore deals with mixtures of refractories and water ice.

\subsubsection{Refractories and $\mathit{H_2O}$} \label{sec_results_hathor_dust_H2O}

Two different types of media consisting of refractories and $\mathrm{H_2O}$ were considered. Both types assume that the 
material once had dust/water--ice mass ratio $\mu=1$, molar abundances $\mathrm{CO_2/H_2O}=0.32$, $\mathrm{CO/H_2O}=0.16$, and
bulk density $\rho_{\rm bulk}=535\,\mathrm{kg\,m^{-3}}$ \citep[for motivations, see][]{davidssonetal22}. The first 
type of models (R14\_006A--R14\_008A) then assumes that the $\mathrm{CO_2}$ and $\mathrm{CO}$ ices have been removed from the 
top region \emph{without any other structural changes to the material}. That produced a dust/water--ice medium with $\psi=0.76$ and 
$\rho_{\rm bulk}=337\,\mathrm{kg\,m^{-3}}$, used for the simulations. In case water ice additionally is removed from the top layer, the dust mantle would have 
$\psi=0.95$ and $\rho_{\rm bulk}=169\,\mathrm{kg\,m^{-3}}$. A Hertz factor ceiling of $h\geq 2.3\cdot 10^{-3}$ was introduced to force a thermal inertia reference value 
$\Gamma\geq 30\,\mathrm{MKS}$ for the dust mantle. The dust/water--ice mixture would have $\Gamma\approx 45\,\mathrm{MKS}$. Because water ice is not visible on most of 
the Hathor alcove surface, one should expect that such a dust mantle indeed has been formed at this location at that time. 

The second type of medium (models R14\_009A--R14\_011A) assumes that the removal of $\mathrm{CO_2}$ and $\mathrm{CO}$ \emph{has led to a compaction 
of the remaining dust/water--ice mixture}. Observational reasons and theoretical mechanisms for such a compaction were primarily discussed by \citet{davidssonetal22b}, but also see \citet{davidssonetal22}. 
Here, such a hypothetical compaction is mainly introduced  to test the influence of the dust mantle heat capacity on the solution. As explained in section~\ref{sec_results_hathor_dust}, 
homogeneous refractories media essentially only have two free parameters: the thermal inertia $\Gamma$ and the ratio $\rho c/E$ (i.~e., $\rho c$ and $E$ cannot be disentangled and 
determined individually). However, when the medium is not homogeneous but layered, and having a significant density discontinuity near the surface, two solutions with different $\rho c$ and $E$ 
are not necessarily identical even when having the same $\rho c/E$ ratio and thermal inertia $\Gamma$. To investigate this effect, the denser media had $\mu=1$ and $\rho_{\rm bulk}=535\,\mathrm{kg\,m^{-3}}$ for the 
dust/water--ice mixture, resulting in $\psi=0.63$. In case a dust mantle forms, it would have a bulk density $\rho_{\rm bulk}=268\,\mathrm{kg\,m^{-3}}$ and porosity $\psi=0.91$. 
In order to achieve a nominal $\Gamma \approx 30\,\mathrm{MKS}$ thermal inertia for the dust mantle, a Hertz factor ceiling of $h\geq 1.4\cdot 10^{-3}$ was used. 

\begin{figure}
\scalebox{0.4}{\includegraphics{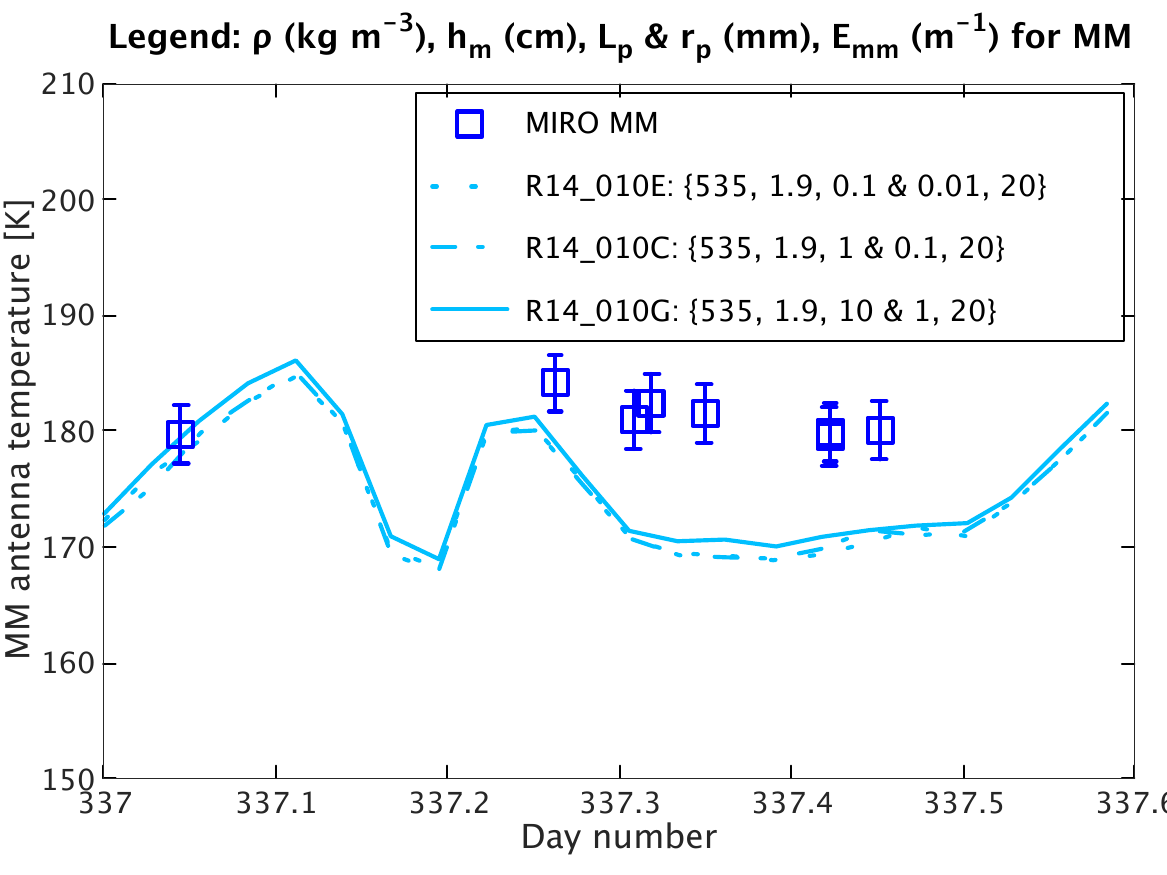}} 
     \caption{Hathor data compared with refractories$+\mathrm{H_2O}$ models with $\Gamma\approx 30\,\mathrm{MKS}$: diffusivity dependence. The (relatively high) dust/water--ice bulk density, mantle thickness, and extinction 
coefficient are held fixed, while the diffusivity is varied, as given in the legend. The insensitivity to diffusivity shows that the water ice is nearly dormant.}
     \label{fig_hathor_diffusivity}
\end{figure}

The first priority, once water ice has been introduced, is to understand the level of sublimation activity and the associated cooling. All models started off with the 
temperature profile of model R14\_001A obtained 28 days prior to the master period and were propagated up to and including that period. No mantle erosion was applied, therefore 
the dust mantle growth directly shows the speed by which water ice withdraws below the surface. Three different values for the diffusivity were applied, based on $\xi=1$ 
and $\{L_{\rm p},\,r_{\rm p}\}=\{100,\,10\}\,\mathrm{\mu m}$, $\{1,\,0.1\}\,\mathrm{mm}$, or $\{10,\,1\}\,\mathrm{mm}$. The low baseline value reproduced the inbound water 
production rate according to \citet{davidssonetal22}. The intermediate and high values corresponds to diffusivities being $\times 10$ and $\times 100$ times higher than the baseline, 
respectively.

A higher diffusivity facilitates escape of vapour from the sublimation front, resulting in a higher net sublimation rate and a higher front withdrawal speed \citep[because we here have weak 
sublimation, whereas the net sublimation rate is insensitive to diffusivity at strong sublimation, see ][]{davidssonetal21}. During the 
28 days, the mantles grew to $0.8$, $1.9$, and $3.2\,\mathrm{cm}$, respectively, for the lower--density models R14\_006A--R14\_008A. The higher--density models 
( R14\_009A--R14\_011A) had $0.4$, $0.8$, and $1.3\,\mathrm{cm}$ thick mantles formed in the same period. Here, the withdrawal is slower because the initial concentration of 
water ice is higher (268 versus $169\,\mathrm{kg\,m^{-3}}$), thus requiring more time for ice removal. 

These mantle formation rates are relatively low, suggesting that water--sublimation cooling may not be particularly efficient under the considered conditions. This suspicion is 
confirmed by Fig.~\ref{fig_hathor_diffusivity}, showing the MM antenna temperature curves for the higher--density models for the three different diffusivities when initial conditions 
have been adjusted to yield the same mantle thickness of $1.9\,\mathrm{cm}$ at the master period. If water sublimation is important, the antenna temperature should be reduced with 
increasing diffusivity because of a more efficient cooling. However, the curves are very similar, i.~e., the diffusivity value has negligible effect on the antenna temperature. The antenna temperature is even 
somewhat elevated at the highest diffusivity, contrary to expectations. This is because a larger tube radius $r_{\rm p}$ also increases the radiative heat transport, here resulting in a slightly 
warmer interior. A similar investigation was made for the lower--density models when the mantle thickness was reduced to $0.8\,\mathrm{cm}$ (thus the degree of sublimation quenching caused by 
the mantle is even weaker). Yet, there was no difference between the low and intermediate diffusivity values, compared to a case where water sublimation was switched off completely by 
artificially forcing the water saturation pressure to zero. From a MIRO observational point of view, the water ice can be considered completely dormant at the Hathor alcove. That is to say, 
while the long--term effect of water sublimation is clearly seen in terms of dust mantle growth, the energy consumption during a given nucleus revolution is too small to have a measurable effect.

\begin{figure}
\scalebox{0.4}{\includegraphics{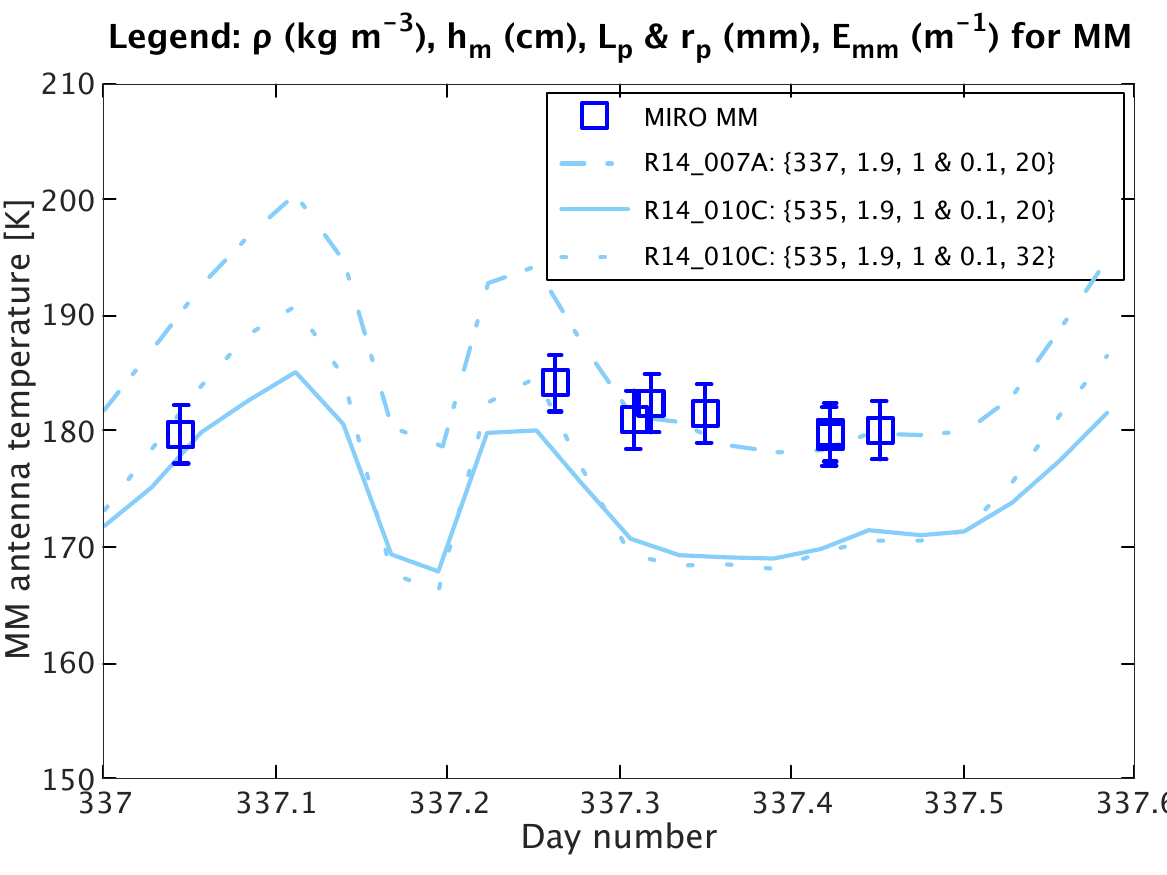}} 
     \caption{Hathor data compared with refractories$+\mathrm{H_2O}$ models with $\Gamma\approx 30\,\mathrm{MKS}$: density (thus heat capacity) dependence. 
Model R14\_007A with mantle density $\rho_{\rm m}=169\,\mathrm{kg\,m^{-3}}$ and extinction coefficient $E_{\rm MM}=20\,\mathrm{m^{-1}}$ and model R14\_010A with 
$\rho_{\rm m}=268\,\mathrm{kg\,m^{-3}}$ and $E_{\rm MM}=32\,\mathrm{m^{-1}}$ have different antenna temperatures. Unlike homogeneous materials, these layered media differ 
despite identical $\rho_{\rm m}/E_{\rm MM}$ ratios. Note that the legend shows the density of the dust/water--ice mixture (the dust mantle values are half as large).}
     \label{fig_hathor_density}
\end{figure}

Next, the influence of the assumed bulk density (and resulting heat capacity) is investigated. Figure~\ref{fig_hathor_density} shows lower--density $\Gamma\approx 30\,\mathrm{MKS}$ 
model R14\_007A that developed a $1.9\,\mathrm{cm}$ mantle after 28 days for the intermediate diffusivity value. At $E_{\rm MM}=20\,\mathrm{m^{-1}}$ the model reproduces the 
nighttime MM data, though the curve amplitude is too high, resulting in a daytime temperature excess. This model has $\rho/E=8.4\,\mathrm{kg\,m^{-2}}$ (here omitting $c$ for simplicity because it 
changes rather slowly with temperature and is similar for all models). The figure also shows higher--density model R14\_010C for the same dust mantle thickness and extinction coefficient 
(resulting in $\rho/E=13.4\,\mathrm{kg\,m^{-2}}$). Here, increasing the mantle density has the effect of lowering the antenna temperature by $11\,\mathrm{K}$ on average, while slightly decreasing 
the amplitude. Both models have similar physical surface temperatures (differing by $4\,\mathrm{K}$ on average). The lower--density model has a skin depth of $L=\Gamma/\rho c\sqrt{\omega}\approx 2.5\,\mathrm{cm}$ 
(for the nucleus rotational angular velocity $\omega=2\upi/P$, where $P$ is the rotational period), which means that the region with strong diurnal temperature variations extends well below the dust mantle. 
The higher--density model has $L\approx 1.5\,\mathrm{cm}$, so that most diurnal temperature variations are confined to the mantle. Because the thermal inertia $\Gamma=\sqrt{\rho c\kappa}$ is fixed, 
the higher--density model R14\_010C has a lower heat conductivity than lower--density model R14\_007A, which is why the former becomes colder at depth. For a homogeneous medium, the R14\_007A solution 
would have been restored by setting $E_{\rm MM}=32\,\mathrm{m^{-1}}$ for model R14\_010C (recovering the $\rho/E=8.4\,\mathrm{kg\,m^{-2}}$ value of model R14\_007A). However, Fig.~\ref{fig_hathor_density} shows that 
the antenna temperature of such a model only recovers partially at day, and remains cold at night. Evidently, the presence of the density discontinuity at shallow depth (caused by the presence 
of inert water ice) distorts the temperature profiles to the point that the simple $\rho c/E$--rule for homogeneous media breaks down. 

For the Hathor alcove (that evidently has dust on its surface), the solutions therefore depend on dust mantle thickness, thermal inertia, \emph{and} mantle bulk density, because it 
is strongly different from the bulk density of the near--surface dust/water--ice mixture. Note that this density--dependence does not apply to the Imhotep dust/water--ice investigation in 
section~\ref{sec_results_Imhotep_dust_H2O} because in that case the dust/water--ice mixture extended up to the surface (i.~e., strong near--surface density gradients are not expected). 
Single--layer media (exemplified by Imhotep) do not allow for unique density solutions (merely identification of the appropriate $\rho/E$ ratio), while two--layer media (exemplified 
by Hathor) allow for constraints to be placed on the density.

\begin{figure*}
\centering
\begin{tabular}{cc}
\scalebox{0.45}{\includegraphics{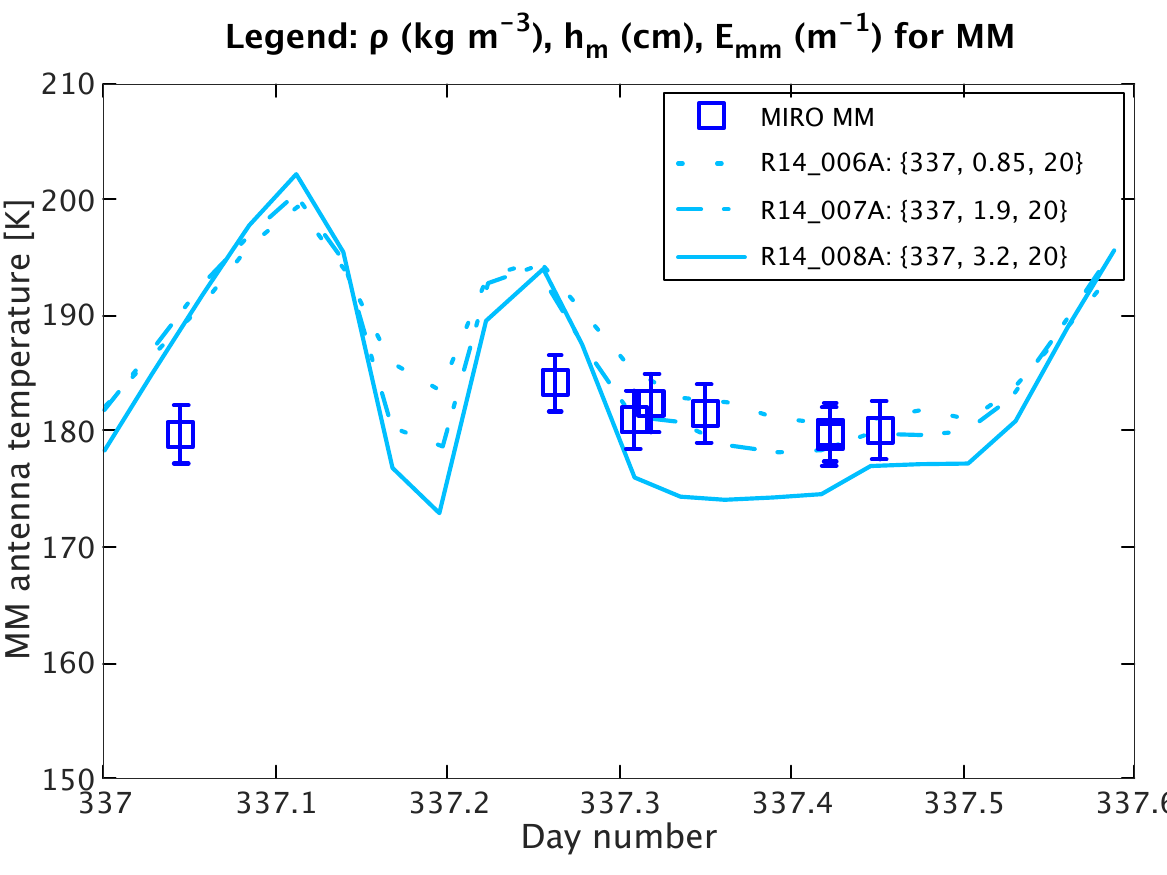}} & \scalebox{0.45}{\includegraphics{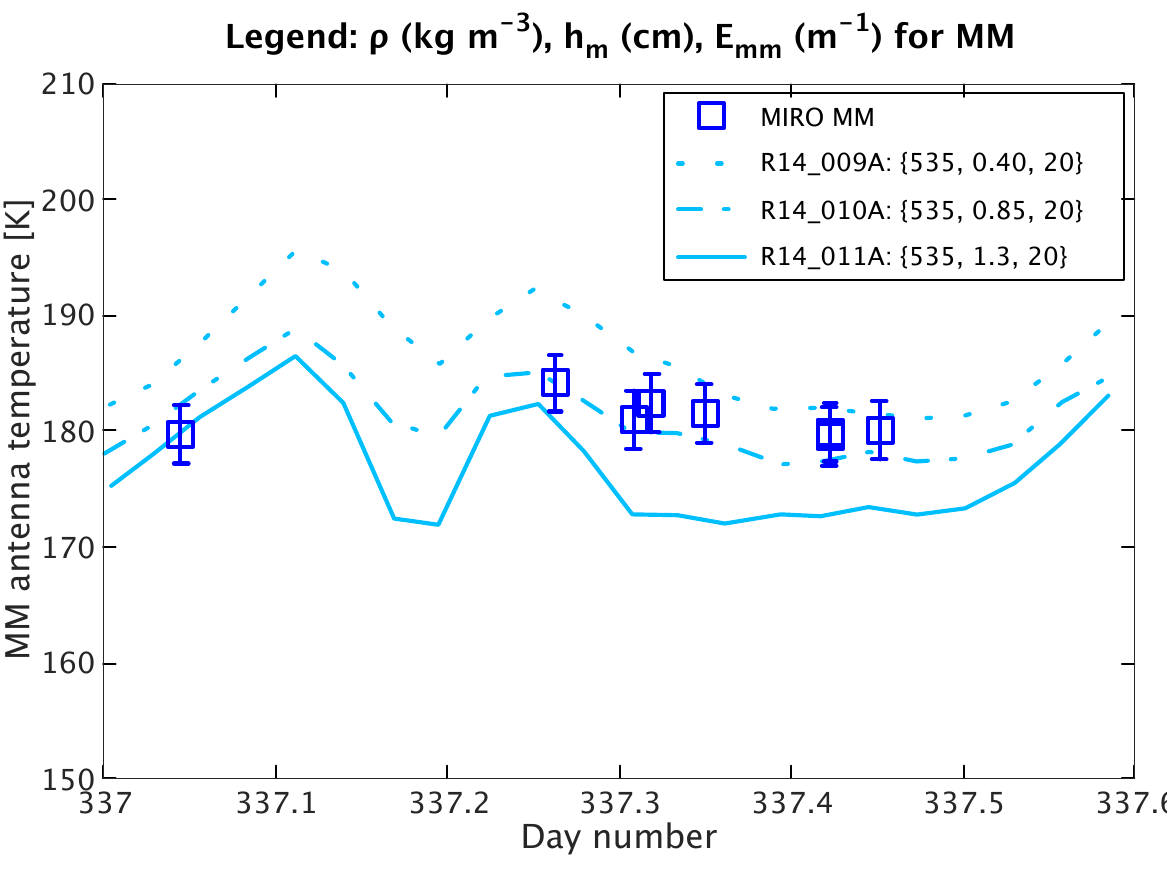}}\\
\end{tabular}
     \caption{Hathor data compared with refractories$+\mathrm{H_2O}$ models with $\Gamma\approx 30\,\mathrm{MKS}$: dust--mantle thickness dependence. 
\emph{Left:} At relatively low densities ($\rho_{\rm m}=169\,\mathrm{kg\,m^{-3}}$ for the mantle, $\rho_{\rm bulk}=337\,\mathrm{kg\,m^{-3}}$ for the dust/water--ice mixture) 
MM fits cannot be obtained at any mantle thickness. \emph{Right:} At relatively high densities ($\rho_{\rm m}=337\,\mathrm{kg\,m^{-3}}$ for the mantle, $\rho=535\,\mathrm{kg\,m^{-3}}$ 
for the dust/water--ice mixture) a MM fit can be obtained for a $0.85\,\mathrm{cm}$ mantle.}
     \label{fig_hathor_comparison}
\end{figure*}

In order to proceed, the MM antenna temperature curves are now studied as function of dust mantle thickness. This is first done for the nominal $\sim 30\,\mathrm{MKS}$ mantle thermal inertia, and 
for the two different mantle density values. That basic comparison in made in Fig.~\ref{fig_hathor_comparison}. The left panel shows the low--density cases. Here, the nighttime data 
can be fitted for $E_{\rm MM}=20\,\mathrm{m^{-1}}$. However, the curve amplitude is high and the models are too warm at day (as previously pointed out, in fact, model R14\_007A appears 
in Fig.~\ref{fig_hathor_density} as well). The nighttime temperature decreases with growing mantle thickness, and for mantle thicknesses $h_{\rm}\stackrel{>}{_{\sim}}  3\,\mathrm{cm}$ it is 
too cold even for an extinction coefficient as low as $E_{\rm MM}=20\,\mathrm{m^{-1}}$. The MIRO MM data cannot be reproduced for a mantle density as low as $\rho_{\rm m}\approx 170\,\mathrm{kg\,m^{-3}}$, 
at least not when $\Gamma\approx 30\,\mathrm{MKS}$.

The right panel of Fig.~\ref{fig_hathor_comparison} shows the corresponding plots for the higher mantle density of $\rho_{\rm m}\approx 340\,\mathrm{kg\,m^{-3}}$. In these cases, the 
curve amplitudes are significantly smaller. Again, the antenna temperature drops with increased mantle thickness. A number of $E_{\rm MM}$--values were considered for each 
mantle thickness, and the best overall fit was obtained for model R14\_010A, having $Q_{\rm MM}=0.08$ for $E_{\rm MM}=20\,\mathrm{m^{-1}}$. Figure~\ref{fig_hathor_comparison} 
shows all models at that extinction coefficient to ease a direct comparison.  Acceptable ($Q_{\rm MM}\geq 0.01$) solutions were only obtained for R14\_010A for the interval 
$E_{\rm MM}=20\pm 10\,\mathrm{m^{-1}}$. 

\begin{figure*}
\centering
\begin{tabular}{cc}
\scalebox{0.45}{\includegraphics{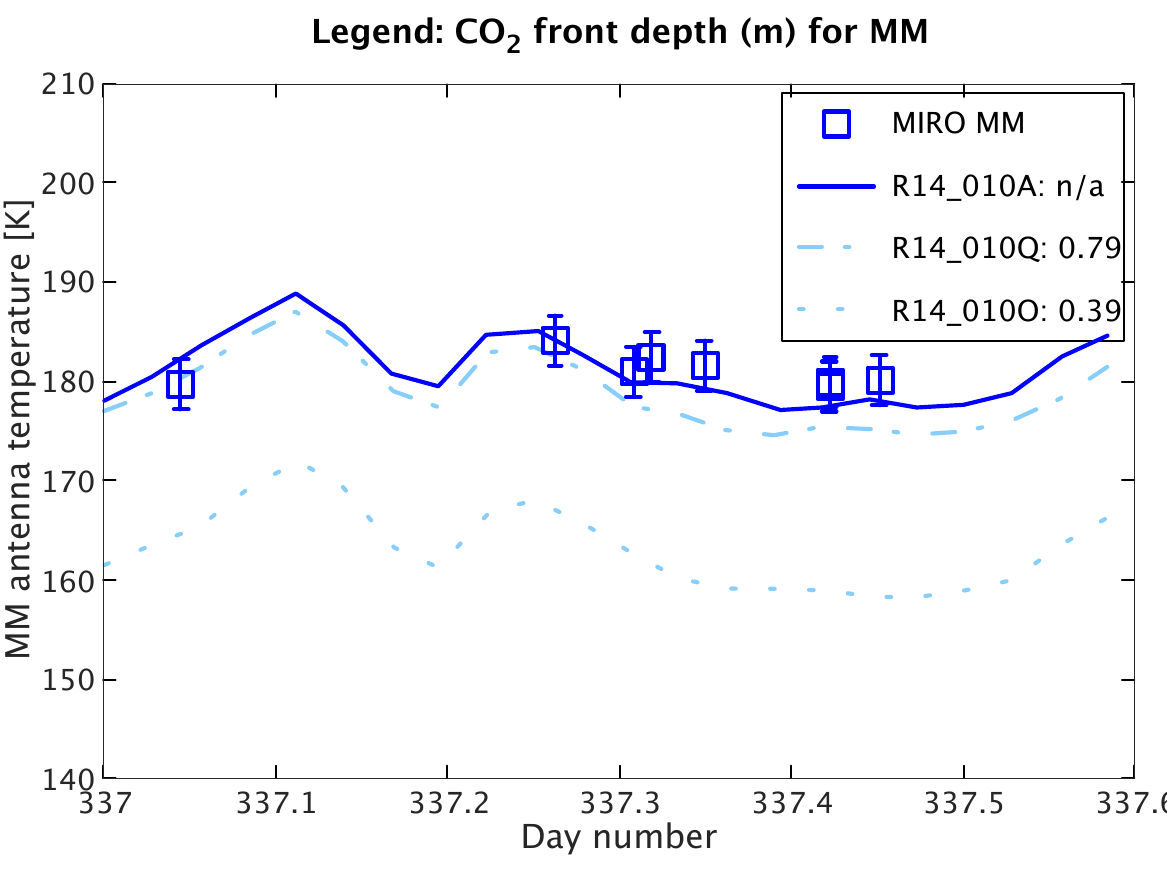}} & \scalebox{0.45}{\includegraphics{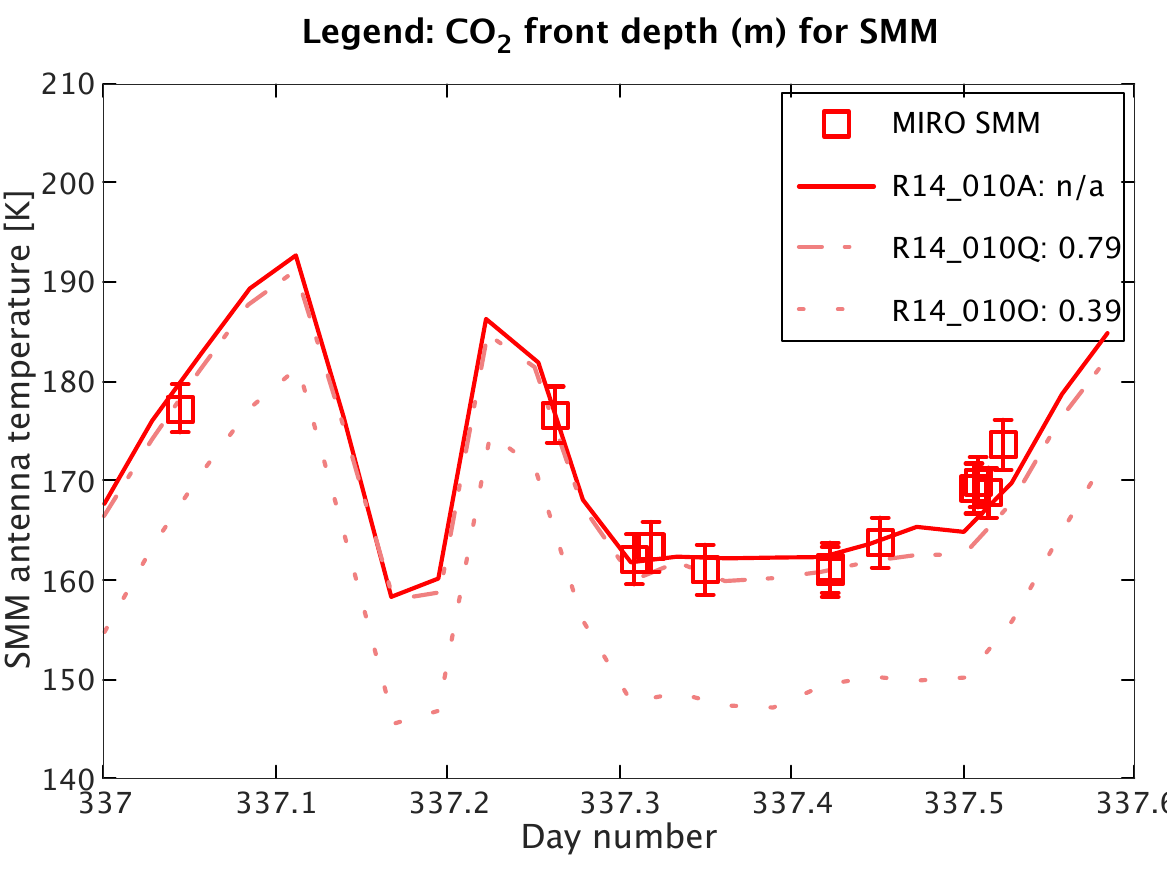}}\\
\end{tabular}
     \caption{Hathor data compared with a refractories$+\mathrm{H_2O}$ model (providing simultaneous MM and SMM fits) and refractories$+\mathrm{H_2O}+\mathrm{CO_2}$ models. 
All models have $h_{\rm m}=0.8\,\mathrm{cm}$ dust mantles with $\rho_{\rm m}=337\,\mathrm{kg\,m^{-3}}$ and $\Gamma\approx 30\,\mathrm{MKS}$, overlaying a dust$+\mathrm{H_2O}$ 
interior with $\rho_{\rm bulk}=535\,\mathrm{kg\,m^{-3}}$. Some models also have dust$+\mathrm{H_2O}+\mathrm{CO_2}$ with $\rho_{\rm bulk}=744\,\mathrm{kg\,m^{-3}}$ below the 
front depths as indicated in the legends (unless not applicable or `n/a'). The $\mathrm{CO_2}$ models have diffusivities based on $\{L_{\rm p},\,r_{\rm p}\}=\{100,\,10\}\,\mathrm{\mu m}$ and $\xi=1$. 
\emph{Left:} MM data and models, all having $E_{\rm MM}=20\,\mathrm{m^{-1}}$ and $w_{\rm MM}=0$. \emph{Right:} SMM data and models with $E_{\rm SMM}=150\,\mathrm{m^{-1}}$ and $w_{\rm SMM}=0.1$.}
     \label{fig_hathor_dust_H2O_CO2}
\end{figure*}

Once one MM solution had been found, attention turned to the SMM data. Among the six $\Gamma\approx 30\,\mathrm{MKS}$ models studied here, the best one was indeed R14\_010A 
for mantle thickness $h_{\rm m}=0.8\,\mathrm{cm}$, with $Q_{\rm SMM}\leq 0.26$ for $E_{\rm SMM}=160\pm 90\,\mathrm{m^{-1}}$, though it was necessary to apply a small level of scattering 
by setting $w_{\rm SMM}=0.10$. Model R14\_011A with $h_{\rm m}=1.3\,\mathrm{cm}$ also had $Q_{\rm SMM}\leq 0.12$ solutions for $E_{\rm SMM}=50\pm 30\,\mathrm{m^{-1}}$ and $w_{\rm SMM}=0$. 
However, a mantle as thick as $h_{\rm m}=1.3\,\mathrm{cm}$ is here excluded for two reasons: 1) there are no corresponding MM solutions; 2) the SMM extinction coefficient is uncharacteristically 
low. Considering those solutions spurious, the conclusion is that at least one simultaneous MM and SMM solution exist for mantles with $\Gamma\approx 30\,\mathrm{MKS}$, $h_{\rm m}=0.8\pm 0.2\,\mathrm{cm}$, $\rho_{\rm m}\approx 340\pm 80\,\mathrm{kg\,m^{-3}}$, $E_{\rm MM}=20\pm 10\,\mathrm{m^{-1}}$, $E_{\rm SMM}=160\pm 90\,\mathrm{m^{-1}}$, and $w_{\rm SMM}=0.1$, overlaying a 
dust/water--ice interior with $\rho_{\rm bulk}\approx 535\,\mathrm{kg\,m^{-3}}$.  The successful MM and SMM model curves are shown in Fig.~\ref{fig_hathor_dust_H2O_CO2} as solid continuous curves.

The next step is to test other mantle thermal inertia values than the nominal $\Gamma\approx 30\,\mathrm{MKS}$. The main questions to be answered are: 1) can solutions be extended to the 
lower--density mantle scenario by modifying the thermal inertia?; 2) can solutions be extended to a wider range of mantle thicknesses in the higher--density mantle scenario by changing the thermal inertia? 
First, four models were run with $\rho_{\rm m}=169\,\mathrm{kg\,m^{-3}}$ for mantle thicknesses $h_{\rm m}=0.8$ and $1.9\,\mathrm{cm}$, considering $\Gamma\approx 15\,\mathrm{MKS}$ or 
$\Gamma\approx 50\,\mathrm{MKS}$ for the mantle. At MM, it was found that increasing the thermal inertia makes the situation somewhat worse, by enhancing the problematic amplitude 
(Fig.~\ref{fig_hathor_comparison}) slightly. Lowering the thermal inertia reduces the amplitude, but the change is so small, that sufficient amplitude reductions for reasonable $\Gamma$ ranges is not deemed possible. 
For that reason, the SMM curves were not considered.

Next, the higher--density case was tested for mantle thicknesses $h_{\rm m}=0.4$, 0.8, and $1.3\,\mathrm{cm}$, for a lower thermal inertia of $15\,\mathrm{MKS}$. This 
destroyed the fit for $h_{\rm m}=0.8\,\mathrm{cm}$ seen at $\Gamma\approx 30\,\mathrm{MKS}$ (by significantly lower the antenna temperature), and it did not introduce new fits at other mantle thicknesses. Therefore, 
attention turned to higher thermal inertia values, first testing $\Gamma=50\,\mathrm{MKS}$ at $h_{\rm m}=0.4$ and $0.8\,\mathrm{cm}$. The changes to the MM curve were very 
marginal for the thinner mantle. However, the quality of the MM fit improved slightly for $h_{\rm m}=0.8\,\mathrm{cm}$ and $\Gamma=50\,\mathrm{MKS}$ compared to the 
lower thermal inertia. The SMM fits were not as good but still acceptable, though the single--scattering albedo had to be increased to $w_{\rm SMM}=0.15$. For this 
reasons, even higher mantle thermal inertia values of $\Gamma=80$ and $100\,\mathrm{MKS}$ were tested for  $h_{\rm m}=0.8\,\mathrm{cm}$. These models were still statistically 
consistent with the data, but reaching amplitudes as low as for R14\_010A with $\Gamma\approx 30\,\mathrm{MKS}$ required $E_{\rm MM}\leq 10\,\mathrm{m^{-1}}$. Such transparent 
media (significant contribution to the MIRO signal from $0.1\,\mathrm{m}$ depth) are probably nonphysical, particularly considering that this is supposed to be a relatively high--density medium. 
The best SMM solutions had $Q_{\rm SMM}\leq 0.02$, thus merely representing marginal fits. Therefore, the thermal inertia range is tentatively restricted to $\Gamma=40\pm 20\,\mathrm{MKS}$.

Note, that increasing $\Gamma$ leads to higher daytime antenna temperatures, while keeping nighttime values virtually fixed (on the $\Gamma=30$--$100\,\mathrm{MKS}$ range). 
Therefore, the low nighttime temperature associated with thicker mantles (i.~e., R14\_011A in Fig.~\ref{fig_hathor_comparison}) are not expected to change much with mantle thermal 
inertia, thus the $h_{\rm m}=1.3\,\mathrm{cm}$ case was not considered. In conclusion, the most likely dust mantle thickness at the Hathor alcove is therefore $h_{\rm m}=0.8\pm 0.2\,\mathrm{cm}$.

\subsubsection{Refractories, $\mathit{H_2O}$, and $\mathit{CO_2}$} \label{sec_results_hathor_dust_H2O_CO2}

Thus far, it has been found that refractories--only models are unable to fit the Hathor alcove data, but that dust/water--ice mixtures 
underneath a thin dust mantle are capable of providing simultaneous MM and SMM fits under the right conditions. For this reason there is 
no strong incentive to introduce $\mathrm{CO_2}$ ice into the study, if the goal is to prove its existence. Even if the presence of $\mathrm{CO_2}$ 
would enable additional simultaneous MM and SMM solutions for other combinations of mantle thicknesses, densities, and thermal inertia values, 
that would not constitute compelling proof that $\mathrm{CO_2}$ necessarily exists near the surface. Clearly, it is possible to do without $\mathrm{CO_2}$, as 
shown in section~\ref{sec_results_hathor_dust_H2O}. 

However,  such fits involving $\mathrm{CO_2}$ (if they are found) could potentially call into question the currently established constraints on 
mantle density, thermal inertia, and thickness. It is therefore important to understand how presence of $\mathrm{CO_2}$ affects the antenna temperature curves. 
For example, the main problem with the lower--density mantles as well as $\Gamma>50\,\mathrm{MKS}$ thermal inertia, was the high amplitudes of the model MM antenna temperature 
curves. If presence of $\mathrm{CO_2}$ would bring down the daytime antenna temperatures without disturbing the nighttime values, that could potentially justify lower--density 
and/or higher--thermal--inertia mantle scenarios. Another reason for considering $\mathrm{CO_2}$ is to understand at what point it would destroy the 
solutions already established in section~\ref{sec_results_hathor_dust_H2O}. Finding the smallest $\mathrm{CO_2}$ front depth for which the current solutions are still acceptable, 
places constraints on how shallow $\mathrm{CO_2}$ possibly can be at the Hathor alcove. 

For this reason, a number of models were run where carbon dioxide ice was introduced up to a given depth. Above the $\mathrm{CO_2}$ front, the parameters of the previously 
successful R14\_010A higher--density model were applied, except that the diffusivity was reduced an order of magnitude, by using $\{L_{\rm p},\,r_{\rm p}\}=\{100,\,10\}\,\mathrm{\mu m}$. 
Below the $\mathrm{CO_2}$ front, $\mathrm{CO_2/H_2O}=0.32$ was applied, resulting in $\rho_{\rm bulk}=744\,\mathrm{kg\,m^{-3}}$ and $\psi=0.49$ at such depths. Figure~\ref{fig_hathor_dust_H2O_CO2} 
shows some of these models. For example, if the $\mathrm{CO_2}$ front is located $\sim 0.4\,\mathrm{m}$ below the surface (model R14\_010O), there are drastic antenna temperature drops both at 
MM (the mean curve is $\sim 17\,\mathrm{K}$ below the mean data) and SMM. If the diffusivity is larger than currently assumed, the discrepancies would grow further. Even when the $\mathrm{CO_2}$ 
front is moved down to $\sim 0.8\,\mathrm{m}$ below the surface (model R14\_010Q), the MM solution is disqualified, although the SMM solution now is compatible with the data. As it turns out, 
the $\mathrm{CO_2}$ front must be located at $\geq 0.99\,\mathrm{m}$ depth to allow for $Q_{\rm MM}\approx 0.01$ solutions (compared to $Q_{\rm MM}=0.08$ for the $\mathrm{CO_2}$--free variant). 

Importantly, Fig.~\ref{fig_hathor_dust_H2O_CO2} shows that the effect of $\mathrm{CO_2}$ sublimation cooling is to lower the entire antenna temperature curve, such that 
the effect on its amplitude is negligible. As a result, $\mathrm{CO_2}$ would not be capable of introducing new solutions at low mantle densities or for high mantle thermal inertia values, 
as discussed previously. Not only does it mean that the constraints placed of mantle density, thermal inertia, and thickness in section~\ref{sec_results_hathor_dust_H2O} are 
confirmed, it also means that $\mathrm{CO_2}$ cannot be allowed to ruin the only solution that has been obtained. That means that the $\geq 1.0\,\mathrm{m}$ lower limit on the 
$\mathrm{CO_2}$ front depth is a necessary criterion.

\section{Discussion} \label{sec_discussion}

The main final conclusions about the estimated conditions at the Imhotep and Hathor collapse sites may be summarised as follows:

\begin{enumerate}
\item \emph{Imhotep:} In 2014 December and 2015 January the MIRO MM and SMM data can be explained by models having $\mu=1$, $\mathrm{CO_2/H_2O}=0.3$, 
$\rho_{\rm bulk}=535\,\mathrm{kg\,m^{-3}}$, $\Gamma=100$--$160\,\mathrm{MKS}$ (with preference for the lower value), $\{L_{\rm p},\,r_{\rm p}\}=\{100,10\}\,\mathrm{\mu m}$, 
$\xi=1$, $E_{\rm MM}=60\pm 10\,\mathrm{m^{-1}}$ (with preference for the upper range), $E_{\rm SMM}=170\pm 40\,\mathrm{m^{-1}}$ (with preference for the lower range), and $w_{\rm SMM}=0$. 
The dust/water--ice mixture is exposed on the surface, and the $\mathrm{CO_2}$ is located at a depth $h_{\rm sv}\approx 11\pm 4\,\mathrm{cm}$ below the surface. 
Here, $\mu$, $\mathrm{CO_2/H_2O}$, $\rho_{\rm bulk}=535\,\mathrm{kg\,m^{-3}}$, $\{L_{\rm p},\,r_{\rm p}\}$, $\xi$ are input parameters based on previous work, while $\Gamma$, $h_{\rm sv}$, 
$E_{\rm MM}$, $E_{\rm SMM}$, and $w_{\rm SMM}$ are estimated in the current paper.
\item \emph{Hathor:} In 2014 November and December the MIRO MM and SMM data can be explained by models having $\mu=1$, $\rho_{\rm m}= 340\pm 80\,\mathrm{kg\,m^{-3}}$, 
$\Gamma=40\pm 20\,\mathrm{MKS}$, $E_{\rm MM}=20\pm 10\,\mathrm{m^{-1}}$, $E_{\rm SMM}=160\pm 90\,\mathrm{m^{-1}}$, and $w_{\rm SMM}=0.1$. The dust mantle has 
a thickness of $h_{\rm m}=0.8\pm 0.2\,\mathrm{cm}$, and the lower limit on the $\mathrm{CO_2}$ front depth is $h_{\rm sv}\geq 1.0\,\mathrm{m}$. All these parameters except $\mu$ are 
estimated in the current work.
\end{enumerate}

These conclusions, along with those for Aswan in Paper~I, allow for an assessment of similarities and differences among three specific nucleus locations. 
First, exposed dust and water--ice mixtures were characterised at Aswan and at Imhotep (both located on the large lobe). Importantly, those are the best available representatives 
of the deep, less processed, nucleus materials. To the extent that abundances, densities, and optical properties can be constrained by the currently applied method, there are no substantial  
differences between interior compositions at Aswan and Imhotep. Both are consistent with dust/water--ice mass ratios near unity, $\mathrm{CO_2/H_2O}$ molar abundances near 30 per cent, 
absence of SMM scattering, and $c\rho/E_{\rm SMM}=2.2\pm 0.2\,\mathrm{kJ\,m^{-2}\,K^{-1}}$. Here, specific heat capacities $c_1=400\,\mathrm{J\,kg^{-1}\,K^{-1}}$ and 
$c_4=1200\,\mathrm{J\,kg^{-1}\,K^{-1}}$ are applied for refractories and water ice, respectively \citep[valid near $T=150\,\mathrm{K}$ according to the laboratory data used by][]{davidsson21}, 
and using the average $800\,\mathrm{J\,kg^{-1}\,K^{-1}}$ for the mixture. The only differences concern thermal inertia (Imhotep having a factor 4--6 higher values than Aswan) and gas diffusivity 
(Imhotep having two orders of magnitude lower values than Aswan). Both may be considered structural properties (heat conductivity, thus thermal inertia, are highly dependent on the level of grain--to--grain 
connectivity, while diffusivity is determined by pore--to--pore connectivity). If these differences are evolutionary, it could be related to the discrepancy in $\mathrm{CO_2}$ sublimation front depths 
at the times of observation: $0.4\pm 0.2\,\mathrm{cm}$ for Aswan but $11\pm 4\,\mathrm{cm}$ for Imhotep. Note that the Aswan diffusivity appears to have dropped to levels similar to those of 
Imhotep as its $\mathrm{CO_2}$ front withdrew to $20\pm 6\,\mathrm{cm}$, indicative of a temporal behaviour that perhaps is shared by most freshly exposed surfaces. The inferred range in 
thermal inertia could easily be accommodated on the rather narrow $0.7\leq\psi\leq 0.8$ interval \citep[see Table~8 in][]{davidsson21} due to the strong dependence of heat conductivity on porosity, 
exemplified by the work of \citet{shoshanyetal02}. If so, the thermal inertia difference may only constitute a temporary fluctuation in the near--surface porosity during a period of intense sublimation. 
Alternatively, these are more deep--rooted (perhaps primordial) differences, that would suggest some level of large--lobe heterogeneity. However, I would caution against such an interpretation until further 
evidence emerges in support of intrinsic variability on the large lobe.

Second, dust--mantle materials were characterised at Aswan on the large lobe (old as well as freshly formed variants) and at Hathor on the small lobe. 
The thermal inertia of Hathor ($\Gamma=40\pm 20\,\mathrm{MKS}$) is similar to the $\Gamma=30\,\mathrm{MKS}$ estimate for pre--collapse Aswan. 
Pre--collapse Aswan had  $c\rho/E_{\rm SMM}=0.2\,\mathrm{kJ\,m^{-2}\,K^{-1}}$ and a substantial level of SMM scattering ($w_{\rm SMM}=0.17$--0.20). The sudden exposure of the 
underlying dust/water--ice mixture led to an order--of--magnitude increase of $c\rho/E_{\rm SMM}$ (to $2.2\pm 0.2\,\mathrm{kJ\,m^{-2}\,K^{-1}}$) and removal of SMM scattering, as mentioned above. 
When the mantle re--formed (though merely being a few millimetres thick) the $c\rho/E_{\rm SMM}$ value dropped (to 0.4 and later to $0.3\,\mathrm{kJ\,m^{-2}\,K^{-1}}$), i.~e., did not quiet reach 
its original level. Notably, SMM scattering was not re--established, perhaps indicating that the mantle needs a certain thickness and/or age to display such properties. The reduction of $c\rho/E_{\rm SMM}$ 
as a consequence of mantle formation is consistent with the removal of water ice, nominally cutting both the specific heat capacity and the bulk density in half. However, this nominally comprises a factor $\sim 4$ 
reduction, while the observed drop is a factor $\sim 11$. The discrepancy could be due to a higher water ice abundance in the real object than inferred here (i.~e., the comet may have $\mu<0.9$), and/or a substantially lower 
specific heat capacity of cometary dust compared to the forsterite model analogue, and/or the result of dust--matrix compaction following the ice removal. The last mechanism would require that the opacity 
increases faster than bulk density during dust--matrix compression (i.~e., that $E_{\rm SMM}$ grows quicker than $\rho$). The Hathor dust mantle had SMM scattering as well 
($w_{\rm SMM}=0.1$) and a comparably large $c\rho/E_{\rm SMM}=1.3\pm 0.7\,\mathrm{kJ\,m^{-2}\,K^{-1}}$. That is a factor $\sim 6$ higher than for the Aswan pre--collapse mantle. 
The MM values are a factor $\sim 5$ higher for Hathor ($c\rho/E_{\rm MM}=9\pm 5\,\mathrm{kJ\,m^{-2}\,K^{-1}}$) than for pre--collapse Aswan ($c\rho/E_{\rm MM}=2.0\pm 0.1\,\mathrm{kJ\,m^{-2}\,K^{-1}}$), 
noting that the re--established Aswan mantle first had $c\rho/E_{\rm MM}=2.3$ and later $1.4\,\mathrm{kJ\,m^{-2}\,K^{-1}}$. This suggests that the 
dust mantle at Hathor is denser, and/or has higher specific heat capacity, and/or is more transparent to microwave radiation than the one at Aswan.

The Aswan/Hathor differences in dust properties may reflect a level of temporal variability that could be common to many nucleus locations, i.~e., they are merely expressions 
of normal fluctuations.  However, the similarity in pre-- and post--collapse Aswan dust mantle $c\rho/E_{\rm SMM}$ and $c\rho/E_{\rm MM}$ values may suggest that 
time variability is not that important. If so, there may be systematic physical and/or chemical differences between Aswan and Hathor, causing their mantles to behave dissimilarly. 
This is an intriguing possibility, considering that Aswan and Hathor are located on different lobes, and remembering that: the small lobe cliffs appear to be stronger than those on the 
large lobe \citep{elmaarryetal16}; polygonal blocks \citep[also referred to as `goosebumps' or `clods', e.~g.][]{davidssonetal16} on the small lobe are twice as large as those on 
the large lobe \citep{fornasieretal21}; bright icy exposures are six times less common on the small lobe as on the large lobe \citep{fornasieretal23}. It is therefore desirable to analyse MIRO data for 
additional locations on both lobes, to establish whether or not the differences inferred here for Aswan and Hathor are systematic.

The Imhotep $\mathrm{CO_2}$ front depth estimate enables an attempt to obtain an approximate age of the feature. This age estimate is based on the results for the 
Aswan collapse site presented in Paper~I. Such a comparison assumes that the differences in thermal inertia and diffusivity discussed above do not 
drastically change the time scale of evolution for freshly exposed interior nucleus material, as long as compositions (using $c\rho/E_{\rm SMM}$ as a proxy) are similar. 
The Aswan structure collapsed on 2015 July 10 \citep[based on OSIRIS observations of a major outburst, see][]{pajolaetal17} and 
the analysis of MIRO observations in Paper~I demonstrated that the $\mathrm{CO_2}$ front was at  $h_{\rm sv}=0.4\pm 0.2\,\mathrm{cm}$ after 5 months, at $h_{\rm sv}=2.0\pm 0.3\,\mathrm{cm}$ 
after 7 months, and at $h_{\rm sv}=20\pm 6\,\mathrm{cm}$ after 11 months. If the Imhotep site has followed the same temporal evolution, the inferred depth of $h_{\rm sv}\approx 11\pm 4\,\mathrm{cm}$ 
in the current work would place the estimated time of collapse $9\pm 1$ months prior to the last days of 2014, or roughly in the 2014 March--May time frame. The OSIRIS cameras 
monitored the brightness of Comet 67P (unresolved at the time) during approach in 2014 March through July, as reported by \citet{tubianaetal15}. Interestingly, the comet displayed a 
larger $\sim 0.6$ magnitude outburst starting 2014 April 27--30, and a smaller $\sim 0.2$ magnitude outburst starting 2014 June 10--20. The rest of the time, the comet magnitude was stable 
to within roughly $\pm 0.05$ magnitudes. Given the similarity of the estimated age of the Imhotep collapse site with the time of the documented larger outburst, I here propose that the larger 
outburst was caused by the collapse that exposed the large patch of dust/water--ice at Imhotep seen in Figs.~\ref{fig_ImhotepA} and  \ref{fig_ImhotepB}. Admittedly, this proposal is speculative 
and impossible to verify.

According to the results for Hathor, the dust mantle has a thickness of $0.8\pm 0.2\,\mathrm{cm}$. This can be compared to the average dust mantle thickness of the northern 
hemisphere of Comet 67P, estimated as $\sim 0.6\,\mathrm{cm}$ based on jet switch--off during sunset according to \citet{shietal16}, and as $\sim 2\,\mathrm{cm}$ based on \textsc{nimbus} model 
reproduction of the empirical water production rate curve of 67P according to \citet{davidssonetal22}. Also, a location in Hapi had a mantle thickness of $h_{\rm m}=2.3\,\mathrm{cm}$ in 2014 October, 
growing to  $h_{\rm m}=21\,\mathrm{cm}$ one month later (possibly due to the presence of a pre--existing ice--free layer) according to the analysis of MIRO data presented by \citet{davidssonetal22b}, 
and the pre--collapse dust mantle thickness at Aswan was  $h_{\rm m}\geq 3\,\mathrm{cm}$ according to Paper~I. In view of these numbers, the recovery of the Hathor dust mantle after a 
presumed collapse in the past is almost or fully complete. Additionally, the necessity of having $\mathrm{CO_2}$ at a depth of $\geq 1.0\,\mathrm{m}$ also suggests that the collapse site in 
Hathor has evolved to a more pronounced level of stratification compared to those in Aswan and Imhotep. For comparison, the average depth of the 
$\mathrm{CO_2}$ sublimation front is $3.8\,\mathrm{m}$ according to \textsc{nimbus} model reproduction of the empirical $\mathrm{CO_2}$ production rate curve of 67P presented by \citet{davidssonetal22}.

Interestingly, the bulk density $\rho_{\rm m}=340\pm 80\,\mathrm{kg\,m^{-3}}$ estimated for the mantle, and the assumed $\rho_{\rm bulk}=535\,\mathrm{kg\,m^{-3}}$ for the 
dust/water--ice mixture (that reproduces data better than for a $\rho_{\rm bulk}=340\,\mathrm{kg\,m^{-3}}$ assumption), are higher than the corresponding values of $\rho_{\rm m}=170$--$190\,\mathrm{kg\,m^{-3}}$ and  $\rho_{\rm bulk}=340$--390$\,\mathrm{kg\,m^{-3}}$ expected if the bulk nucleus with a measured $\rho_{\rm bulk}=535\,\mathrm{kg\,m^{-3}}$ simply is deprived of its $\mathrm{CO}$ and $\mathrm{CO_2}$. 
The higher values suggest compression from porosities $\psi=0.76$ to $\psi=0.63$ for the dust/water--ice mixture, and from $\psi=0.95$ to $\psi=0.91$ for the dust mantle. A major outcome of the 
\emph{Rosetta} mission, based on measurements by CONSERT and SESAME--PP, is that the near--surface region is compressed compared to the deeper interior, particularly in the upper metre
 \citep{ciarlettietal15,ciarlettietal18,brouetetal16,lethuillieretal16}. Prior to \emph{Rosetta}, \citet{davidssonetal09} also proposed the existence of a near--surface compacted layer based on 
discrepancies in density estimates from radar (valid for the upper few metres) and from non--gravitational force modelling (valid for the bulk nucleus). The current results for Hathor is consistent with 
that global property. \citet{davidssonetal22b} has proposed that the $\mathrm{CO_2}$ vapour pressure profile, that falls steeply on both sides of the $\mathrm{CO_2}$ front 
(according to \textsc{nimbus} simulations) is responsible, by forcing movements in the solid dust/water--ice matrix \citep[also see][]{davidssonetal22}. 
The inward gradient below the front is hypothesised to cause compression that is considered the ultimate reason of the observed near--surface compaction. The outward pressure gradient above the front (of both 
$\mathrm{H_2O}$ and $\mathrm{CO_2}$ vapour) has traditionally been considered the cause of ejection of material into the coma \citep[e.g.,][]{fanaleandsalvail84}, so it is obvious that the vapour is capable of 
forcing these structural changes.

It is perhaps significant that the SMM observations of Hathor seem to require a certain level of scattering at such wavelengths. A similar single--scattering albedo ($w_{\rm SMM}$) was obtained 
for a location in Hapi by \citet{davidssonetal22c}, and seemed to be related to a compaction event of the surface that additionally increased the thermal inertia, MM and SMM extinction coefficients, 
reduced the diffusivity, and removed a previously detectable solid--state greenhouse effect. Scattering is believed to be caused by small--scale dielectric constant variations \citep{garykeihm78}, caused by 
millimetre--scale lumpiness. Such lumpiness would form if an initially uniform mixture of $\mathrm{\mu m}$--sized monomer grains is compressed in such a manner that millimetre--sized regions of higher compaction 
are mixed with lower--density regions of similar size. The emergence of SMM scattering after compression of cometary material, may provide clues on how the small--scale structure of such media are 
modified when compressed.

The analysis of the Imhotep collapse site resulted in an ambiguity -- whereas solutions with dust and water ice could be excluded, the MIRO data was not sufficient to 
distinguish between (nearly) equally good solutions for refractories--only and refractories mixed with both $\mathrm{CO_2}$ and $\mathrm{H_2O}$. The reason for this 
difficulty was the quality of the antenna temperature curves, that lacked data for rotational phases that would have allowed for a disentanglement. Had it not been for the 
OSIRIS images that exclude dust coverage at the Imhotep site, it would have been impossible to select the dust/$\mathrm{H2O}$/$\mathrm{CO_2}$ solution as the correct one. 
Based on this difficulty, and similar ones described in \citet{davidssonetal22b} and Paper~I, the following operational and observational recommendations are being made for 
future spacecraft missions carrying a microwave instrument:

\begin{enumerate}
\item Close collaborations between the camera and microwave instrument teams are necessary. The camera team will identify regions of particular interest (e.~g., cliff collapses) 
in real time, but it is crucial that such information is passed on to the microwave instrument team and others, to enable urgent follow--up observations with other types of instrumentation. 
It also requires a certain level of flexibility in spacecraft operations, i.~e., to allocate time for target--of--opportunity observations, which is not possible when observations are planed 
in month--long blocks, weeks in advance, and executed without regard to developing events. 
\item It is crucially important that a microwave instrument observes a given location during large fractions of the comet day and night, in order to cover as many rotational phase angles as possible. Only 
with access to a complete thermal diurnal antenna temperature curve is it possible to determine the thermophysical parameters of the location with high accuracy and without ambiguities. It is far 
better to observe a limited number of carefully selected regions repeatedly, than to continuously scan the nucleus in an attempt to cover as much ground as possible.
\item The attempt to analyse microwave data has revealed the importance of access to observations at different wavelength intervals. Sometimes, cases can be found where one physical model 
of the nucleus reproduces the MM measurements, whereas another physical model of the nucleus reproduces the SMM measurements (e.~g., section~3.3 in Paper~I). Obviously, none of those 
physical models are correct, and a third physical model that simultaneously satisfies both the MM and SMM observations needs to be found. Otherwise, false solutions risk to confuse and distort 
the description of the comet nucleus.  
\end{enumerate}

Finally, I echo the conclusion of Paper~I, that newly collapsed cliff sites may be the best option for a cryogenic comet nucleus sample return mission that aims at finding super--volatiles like 
$\mathrm{CO_2}$ in addition to the more readily available water ice. Based on the analysis of Aswan in Paper~I, and of Imhotep in the current work, it is found that 
$\mathrm{CO_2}$ ice manages to stay close (centimetres to decimetres) to within the surface for up to a year after collapse. Extracting ice from such depths requires less technically complex 
and less costly engineering solutions, than targeting locations elsewhere that may require drilling through several metres of material. It is primarily the long--lived exposures of dusty water ice 
that reveals the presence of shallow $\mathrm{CO_2}$. Combined with the rarity of exposed water ice elsewhere on the comet, Paper~I and the current work suggest that it is 
the strong near--surface $\mathrm{CO_2}$ sublimation that prevents the dust mantle from forming, whereas water ice is incapable of self--cleaning.

\section{Conclusions} \label{sec_conclusions}

This paper has analysed observations by \emph{Rosetta}/MIRO of Comet 67P/Churyumov--Gerasimenko with the aid of the thermophysical 
model \textsc{nimbus} \citep{davidsson21} and radiative transfer model \textsc{themis} \citep{davidssonetal22b}, with the goal of 
constraining the physical properties (including thermal inertia, density, dust mantle thickness, $\mathrm{CO_2}$ sublimation front depth, extinction 
coefficients, and single--scattering albedos) of near--surface comet nucleus material at two cliff collapse sites -- the relatively fresh Imhotep location, 
and the relatively aged Hathor alcove. The main results are summarised as follows:

\begin{enumerate}
\item The MIRO observations of the Imhotep collapse site (an exposed dust/water--ice mixture) are consistent with a dust/water--ice mass ratio $\mu=1$, a molar $\mathrm{CO_2}$ abundance 
relative to water of $\mathrm{CO_2/H_2O}=0.3$, an interior bulk density $\rho_{\rm bulk}=535\,\mathrm{kg\,m^{-3}}$, a thermal inertia $\Gamma=100$--$160\,\mathrm{MKS}$ 
(with preference for the lower value), a gas diffusivity parameterised by $\{L_{\rm p},\,r_{\rm p}\}=\{100,10\}\,\mathrm{\mu m}$ and $\xi=1$, extinction coefficients 
$E_{\rm MM}=60\pm 10\,\mathrm{m^{-1}}$ (with preference for the upper range) and $E_{\rm SMM}=170\pm 40\,\mathrm{m^{-1}}$ (with preference for the lower range), and single--scattering 
albedos $w_{\rm MM}=w_{\rm SMM}=0$. The $\mathrm{CO_2}$ ice is located at a depth $h_{\rm sv}\approx 11\pm 4\,\mathrm{cm}$ below the surface. 
\item The estimated age of the Imhotep site ($9\pm 1$ months at the time of observation) is consistent with an outburst observed by \emph{Rosetta}/OSIRIS during 
approach to Comet 67P, here proposed to pinpoint the time of collapse to 2014 April 27--30.
\item The MIRO observations of the Hathor alcove (primarily covered with dust but containing patches of exposed water ice) are consistent with $\mu=1$, 
$\rho_{\rm m}= 340\pm 80\,\mathrm{kg\,m^{-3}}$, $\Gamma=40\pm 20\,\mathrm{MKS}$, $E_{\rm MM}=20\pm 10\,\mathrm{m^{-1}}$, $E_{\rm SMM}=160\pm 90\,\mathrm{m^{-1}}$, $w_{\rm MM}=0$, 
and $w_{\rm SMM}=0.1$. The dust mantle has a thickness of $h_{\rm m}=0.8\pm 0.2\,\mathrm{cm}$, and the lower limit on the $\mathrm{CO_2}$ front depth is $h_{\rm sv}\geq 1.0\,\mathrm{m}$.
\item The increased dust mantle density ($\rho_{\rm m}= 340\pm 80\,\mathrm{kg\,m^{-3}}$ compared to the theoretically expected $\rho_{\rm m}=170$--$190\,\mathrm{kg\,m^{-3}}$) suggests 
near--surface compaction, which is consistent with \emph{Rosetta}/CONSERT and SESAME--PP measurements.  
\end{enumerate}

\section*{Acknowledgements}
The author appreciates the useful recommendations and thoughtful comments of the reviewer, Dr. M. Ramy El--Maarry. The author is indebted to 
Dr. Pedro J. Guti\'{e}rrez from Instituto de Astrof\'{i}sica de Andaluc\'{i}a, Granada, Spain, for invaluable discussions and comments. 
The research was carried out at the Jet Propulsion Laboratory, California Institute of Technology, under a contract with the National Aeronautics and Space Administration. 
The author acknowledges funding from NASA grant 80NM0018F0612 awarded by the \emph{Rosetta} Data Analysis Program.\\

\noindent
\emph{COPYRIGHT}.  \textcopyright\,2024. California Institute of Technology. Government sponsorship acknowledged.

\section*{Data Availability}

The data underlying this article will be shared on reasonable request to the corresponding author.

\bibliography{MN-23-3753-MJ.R1.bbl}

\bsp	
\label{lastpage}
\end{document}